\documentclass[twocolumn,twocolappendix]{aastex63}

\received{date}
\revised{date}
\accepted{\today}

\shorttitle{The CGM of BreakBRD Galaxies}
\shortauthors{Tonnesen, DeFelippis \& Tuttle}

\graphicspath{{./}{figures/}}

\begin{document}

\title{
You Are What You Eat: The Circumgalactic Medium Around BreakBRD Galaxies has Low Mass and Angular Momentum }

\author[0000-0002-8710-9206]{Stephanie Tonnesen}
\affiliation{CCA, Flatiron Institute, 162 5th Avenue, New York, NY 10010, USA}

\author[0000-0002-0112-7690]{Daniel DeFelippis}
\affiliation{Universit\'e Paris-Saclay, Universit\'e Paris Cit\'e, CEA, CNRS, AIM, 91191, Gif-sur-Yvette, France}

\author[0000-0002-7327-565X]{Sarah Tuttle}
\affiliation{University of Washington, Seattle, Physics and Astronomy Building, Seattle, WA, 98195, USA}

\correspondingauthor{Stephanie Tonnesen}
\email{stonnesen@flatironinstitute.org}

%

\begin{abstract}

Observed breakBRD (“break bulges in red disks") galaxies are a nearby sample of face-on
disk galaxies with particularly centrally-concentrated star formation: they have red disks
but recent star formation in their centers as measured by the D$_n$4000 spectral index. In Kopenhafer et al. (2020), a comparable population of breakBRD analogues was identified in the TNG simulation, in which the central concentration of star formation was found to reflect a central concentration of dense, star-forming gas caused by a lack of dense gas in the galaxy outskirts.  In this paper we examine the circumgalactic medium of the central breakBRD analogues to determine if the extended halo gas also shows differences from that around comparison galaxies with comparable stellar mass.  We examine the circumgalactic medium gas mass, specific angular momentum, and metallicity in these galaxy populations.  We find less gas in the circumgalactic medium of breakBRD galaxies, and that the breakBRD circumgalactic medium is slightly more concentrated than that of comparable M$_*$ galaxies.  In addition, we find that the angular momentum in the circumgalactic medium of breakBRD galaxies tends to be low for their stellar mass, and show more misalignment to the angular momentum vector of the stellar disk.  Finally, we find that the circumgalactic medium metallicity of breakBRD galaxies tends to be high for their stellar mass.  Together with their low SFR, we argue that these CGM properties indicate a small amount of disk feeding concentrated in the central regions, and a lack of low-metallicity gas accretion from the intergalactic medium.

\end{abstract}



\section{Introduction} \label{sec:intro}

The circumgalactic medium (CGM), halo gas that contains roughly half of the baryonic mass in dark matter halos \citep{Werk_2014ApJ...792....8W, TUmlinson_2017ARA&A..55..389T}, is now understood to be a critical component of the gas ecosystem that regulates galaxy evolution. It is the environment through which cosmological accretion must flow to reach the galaxy disk, and into which mass, momentum and energy are injected from either stellar or black hole feedback.  Due to the gas cycling between the galaxy and the CGM, we expect the state of the CGM to be connected to galaxy properties, particularly for central (defined here as galaxies which are not identified as satellites, so analogous to the Milky Way as a central galaxy) galaxies whose CGM is less impacted by the surrounding environment.  

Because gas flowing from the CGM feeds star formation in galaxy disks, and supernovae feedback ejects mass and metals back into the CGM, observers have looked for correlations between the interstellar medium (ISM) cold gas content or star formation rate (SFR) of galaxies and the amount of cold gas in the CGM.  Indeed, COS-GASS has found that the amount of cold gas in the ISM and CGM of galaxies is positively correlated \citep{Borthakur_2015ApJ...813...46B}.  In addition, cold gas traced by Mg II has been found to positively correlate with galaxy SFR, with larger Mg II equivalent widths around blue galaxies than around red galaxies \citep{Bordoloi_2011ApJ...743...10B}, and larger Mg II covering fractions around star-forming galaxies compared to quiescent galaxies  \citep{Huang_MgII_2021MNRAS.502.4743H}.  However, cold gas traced by HI tells a less consistent story.  While COS-Halos found very little difference in the amount of HI in the CGM of star-forming versus passive galaxies \citep{Tumlinson_COSHalos_2013ApJ...777...59T}, an SDSS selected sample of starburst and post-starburst galaxies observed stronger HI at large radius compared with the COS-Halos and COS-GASS samples containing galaxies with lower SFRs \citep{Heckman_COSBurst_2017ApJ...846..151H}.  

 In addition to less cold gas generally being found around quiescent galaxies, O VI appears to be absent around the non-star-forming, more massive galaxies in the COS-Halos sample \citep{Tumlinson_OVI_2011Sci...334..948T}, and is found in excess around late-type galaxies \citep[see also][]{Johnson_excesscoolgas_2015MNRAS.452.2553J, Chen_Mulchaey_2009ApJ...701.1219C}.  The dearth of $\sim$10$^{5.5}$ K gas around quenched galaxies could indicate either a hotter CGM or non-equilibrium cooling \citep{Oppenheimer_2016MNRAS.460.2157O}.  

Star formation feedback may also have an effect on the spatial distribution of absorbers within the CGM.  Several groups have observed that Mg II and O VI absorption tends to lie along the minor galaxy axis in star-forming galaxies, indicating feedback-driven outflows are correlated with cooler CGM gas \citep{Bordoloi_2011ApJ...743...10B, Bouche_2012MNRAS.426..801B, Kacprzak_2012ApJ...760L...7K, Kacprzak_2015ApJ...815...22K,Nielsen_2015ApJ...812...83N,Schroetter_MEGAFLOW_2019MNRAS.490.4368S, Martin_Ho_CGMkinematics_2019ApJ...878...84M}.  Low-metallicity gas inflowing from the intergalactic medium through the CGM tends to be observed along a galaxy's major axis \citep[e.g.][]{Crighton_2013ApJ...776L..18C,Nielsen_2015ApJ...812...83N}, corotating with the galactic disk, and possibly forming thick rotating ``disks'' of halo gas before inspiraling \citep{Steidel_2002ApJ...570..526S,Ho_2017ApJ...835..267H,DiamondStanic_2016ApJ...824...24D}.  Such cold extended disks in the CGM have also been found in zoom-in simulations run with several different hydrodynamical codes \citep{Stewart_2013ApJ...769...74S, Stewart_2017ApJ...843...47S}.

Simulations also find that the CGM is strongly connected to  galaxy evolution via the gas cycle. IGM gas accretion through the CGM along the disk plane adds high angular momentum, metal-poor gas to fuel star formation \citep[e.g.][]{Dekel_2009Natur.457..451D,Dekel_2013MNRAS.435..999D,vandevoort2011MNRAS.414.2458V,Brook2011MNRAS.415.1051B,Stewart_2011ApJ...738...39S,Brook_2012MNRAS.419..771B,Ubler_2014MNRAS.443.2092U,Christensen_2016ApJ...824...57C,DeFelippis_2017ApJ...841...16D,Grand_Auriga_2019MNRAS.490.4786G}.  Concurrently, gas expelled from the galaxy via feedback flows along the minor axis with low angular momentum \citep{Brook2011MNRAS.415.1051B,Brook_2012MNRAS.419..771B,Ubler_2014MNRAS.443.2092U,Mitchell_EAGLErecycling_2020MNRAS.497.4495M,DeFelippis2020ApJ...895...17D}.  The gas along the minor axis then has higher metallicity as well as lower angular momentum than that aligned with the disk plane, thus far measured in the TNG cosmological suite of simulations \citep{Peroux_TNG_2020MNRAS.499.2462P, Truong_2021MNRAS.508.1563T}.  We note that although the sample sizes are currently small, observations so far do not find the strong trend of metallicity varying with azimuthal angle identified in simulations \citep{Peroux_2016MNRAS.457..903P,Kacprzak_2019ApJ...886...91K,Pointon_2019ApJ...883...78P}. 

In broad agreement with observations, simulations have found that the CGM systematically varies between star-forming and quenched galaxies. Recently, much of this work has been done using the TNG simulation suite that we examine in this paper \citep{Marinacci18,Naiman18,Nelson18,Pillepich18,Springel18}.  For example, using the TNG100 and TNG300 simulations, \citet{Nelson_TNGoxygen_2018MNRAS.477..450N} found more O VI mass around star-forming galaxies compared to quenched galaxies of the same mass.  \citet{Fielding_SMAUG_2020ApJ...903...32F} found that in TNG100 quenched galaxies had hotter median temperatures in their inner CGM, but higher cold gas fractions in their outskirts.  However, in agreement with previous work using the EAGLE simulations, quenched galaxies had lower total CGM mass \citep{Davies_2019MNRAS.485.3783D,Davies_EAGLECGM_2020MNRAS.491.4462D}, so did not necessarily have higher amounts of cold gas in comparison to the CGM of star-forming galaxies. Studying galaxies in the TNG100 simulations,  \citet{DeFelippis2020ApJ...895...17D} (D20) found that the angular momentum of the CGM is both larger and more aligned with the disk angular momentum in galaxies with high stellar angular momentum (high stellar angular momentum has been found to correlate with high SFR by \citet{genelGalacticAngularMomentum2015}). Interestingly, other authors have found that in the same simulation suite (TNG100), the angular momentum in the CGM is higher around quenched galaxies \citep{Wang_TNGAM_2022MNRAS.509.3148W, Lu_TNGAM_2022MNRAS.509.2707L}, although this result is for somewhat more massive galaxies than D20.

While both observations and simulations are continuing to find connections between central galaxies and their surrounding CGM, it is still not clear how directly the state of the CGM reflects the gas and SF distribution in the galaxy disk.  In this paper we search for this direct reflection by studying the CGM around a small sample of galaxies with unusual gas and SF distributions: Break Bulge, Red Disk galaxies (breakBRD), galaxies near $z$ = 0 observationally identified to have centrally-concentrated SF using the D$_n$4000 break and red surrounding disks using ($g - r$) color \citep{tuttleBreakBRDGalaxiesGlobal2020}.  

\citet{Kopenhafer2020ApJ...903..143K} identified breakBRD analogues in TNG100 \citep{Marinacci18,Naiman18,Nelson18,Pillepich18,Springel18}, and found that the simulated analogues had a dearth of star-forming gas outside their central 2 kpc in addition to a dearth of star formation.  Their central gas content and SFR was similar to a stellar mass matched control sample, resulting in somewhat low global gas content and SFR, although they are not quenched.  However, by tracking a breakBRD analogue sample identified at $z$ = 0.5, K20 found that breakBRD galaxies are more likely to quench than galaxies in a mass-matched sample.

Because of their unusually centrally-concentrated and globally low, but not quenched, SFR, it is interesting to examine the CGM of breakBRD galaxies.  In this paper we focus on the CGM mass, angular momentum, and metallicity, all features of the CGM that have been shown to correlate with properties of central galaxies in cosmological simulations.  We will determine if the CGM of breakBRDs is different from galaxies of similar stellar mass, with the dual goals of making predictions for future CGM observations of the observed breakBRD sample and completing our understanding of the breakBRD gas cycle by connecting the CGM gas to the disk gas.  

The paper is organized as follows. Section \ref{sec:method} first briefly introduces the TNG simulations (Section \ref{sec:TNG100}), discusses the breakBRD and comparison sample selections used in this paper (Section \ref{sec:method_sample}), and finally defines our cold CGM criteria (Section \ref{sec:method_coldgas}).  We present our global CGM measures in Section \ref{sec:global}.  In Section \ref{sec:maps} we examine the CGM in spatial detail using maps of the mass, angular momentum, and metallicity distribution around breakBRD galaxies and the comparison sample.  We discuss our results with regards to the gas cycle in breakBRD galaxies and the CGM-galaxy connection in Sections \ref{sec:discussion_gascycle} \& \ref{sec:discuss_trends}, and make observational predictions in Section \ref{sec:discussion_obs}.  Finally, we summarize our conclusions in Section \ref{sec:conclusions}.

\section{Method} \label{sec:method}

\subsection{TNG100}\label{sec:TNG100}
The IllustrisTNG100 simulation \citep[public data release: ][]{Marinacci18,Naiman18,Nelson18,Pillepich18,Springel18}\footnote{www.tng-project.org} is part of a suite of cosmological simulations run using the AREPO moving mesh code \citep{Springel2010}.  TNG100 has a volume of  110.7 Mpc$^3$ and a mass resolution of $7.5 \times 10^6 \mathrm{M}_{\odot}$ and $1.4 \times 10^6 \mathrm{M}_{\odot}$ for dark matter and baryons, respectively.  The TNG suite implements upgraded subgrid models compared to the Illustris simulation \citep{vogelsbergerIntroducingIllustrisProject2014, Genel2014}; specifically, a modified black hole accretion and feedback model \citep{Weinberger2017}, updated galactic winds \citep{Pillepich2018}, as well as the addition of magnetohydrodynamics \citep{Pakmor2011}, all resulting in more realistic galaxies compared to the original Illustris simulation \citep[e.g.][]{Nelson18}.

\subsection{The BreakBRD and comparison sample selection}\label{sec:method_sample}

As introduced above, breakBRD galaxies were first found in SDSS as unusual nearby (z $<$ 0.05) galaxies that have star-forming central regions (using the D$_n$4000 break) embedded in red disks (using ($g - r$) colors) \citep{tuttleBreakBRDGalaxiesGlobal2020}.  The BreakBRD analogue sample of galaxies was selected from within the TNG cosmological simulation, and is defined in \citet{Kopenhafer2020ApJ...903..143K}, but here we briefly summarize the main selection criteria.

Our first criterion was that the subhalo stellar mass must lie within $10^{10} < M_\ast < 10^{12}\ \mathrm{M_{\odot}}$. We chose our lower mass limit in order to resolve galaxies and their central 2~kpc since our analysis requires looking directly at the central region of galaxies.  We also ignored galaxies with $M_* > 10^{12} \mathrm{\ M_\odot}$, as this is outside the mass range of the observed breakBRD sample, and we assumed these galaxies were mainly ellipticals.  

We required galaxies to have $R_{0.5M} > 2$~kpc, where $R_{0.5M}$ is the stellar half mass radius. This requirement removed galaxies which did not have a well-resolved difference between the central region and the outskirts, and therefore would not be meaningful additions to our sample. 

To select our sample of breakBRD analogues from TNG, we then calculated star formation histories of all the galaxies in our mass- and size-defined parent sample. We use their star formation histories and the Flexible Stellar Population Synthesis (FSPS) code of \cite{conroyPropagationUncertaintiesStellar2009} \citep[updated in][]{conroyPropagationUncertaintiesStellar2010}, with the Python interface from \citet{danforeman-mackeyPythonfspsPythonBindings2014}, to generate mock spectra for the inner $r < 2$~kpc region, and \textit{g} and \textit{r} colors for the disk region ($2\mathrm{\ kpc} < r < 2 R_{0.5M}$) of our parent sample. 

To calculate the D$_n$4000 measure we apply the narrow definition from \citet{baloghDifferentialGalaxyEvolution1999}. Galaxies with $\mathrm{D_n4000} < 1.4$ in the inner 2~kpc comprise the D$_n$4000 selection. We also select galaxies with red outskirts using a color cut of $g-r > 0.655$ in the $2\mathrm{\ kpc} < r < 2 R_{0.5M}$ region. 

Our final sample consists of galaxies that exhibit both a D$_n$4000 break in the bulge and a red disk, consisting of 235  galaxies at redshift $z=0.0$ (out of 6092 mass- and size-selected galaxies). This sample is the complete breakBRD analogue sample.

In this paper, we focus on the central galaxies in the breakBRD analogue sample, a subset of 88 galaxies.  Throughout the paper, `breakBRD galaxies' refers to the subset of central breakBRD analogue galaxies (this is labeled BBRD in some figures).  One of these galaxies has an abnormally large stellar mass of $\approx 2\times10^{11} \; M_{\odot}$ compared to the rest of the sample, so we exclude it from most of this analysis. The majority of the other 87 galaxies have stellar masses between 10$^{10}$ - 10$^{10.5}$ M$_{\odot}$, so for the detailed comparisons between the CGM of breakBRD galaxies and the general population in TNG in Section \ref{sec:maps} our comparison sample consists of all central galaxies within this stellar mass range.

\begin{figure}
    \centering
    \includegraphics[scale=0.61]{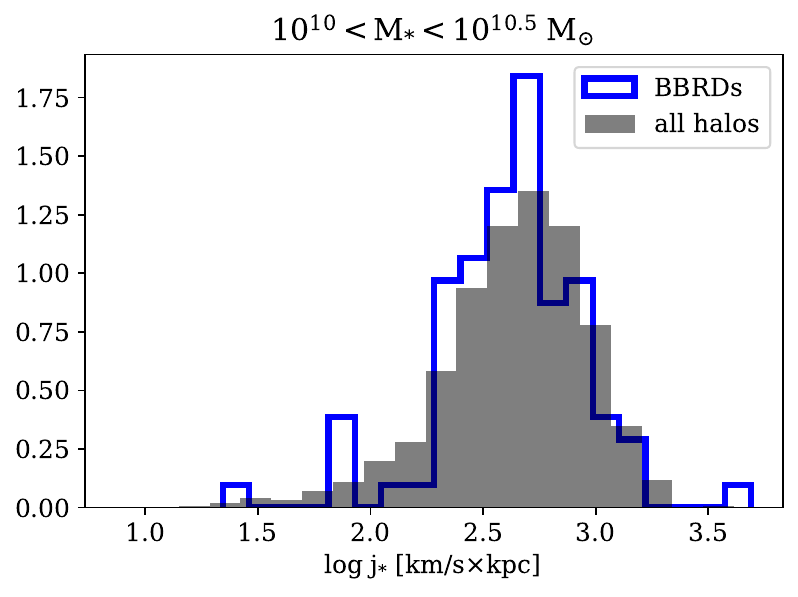}
    \caption{Normalized histogram of j$_*$ of central BBRD analogue galaxies (88 galaxies, blue histogram) and all central halos with stellar masses between 10$^{10}$ - 10$^{10.5}$ M$_{\odot}$ (2132 galaxies, grey histogram).  While the BBRD analogue population median j$_*$ is somewhat lower than that of the comparison sample, a KS test has a p-value of 0.32, indicating that the distributions are not significantly different from each other. }
    \label{fig:jstarhist}
\end{figure}

We also split our galaxies into those with low- and high-j$_*$, where j$_*$ is defined as the specific angular momentum of all of the stellar particles belonging to the galaxy (the same definition used in \cite{DeFelippis2020ApJ...895...17D}).
As shown in \citet{genelGalacticAngularMomentum2015}, j$_*$ is correlated with many galaxy properties, including star formation rate, where it is seen that quiescent galaxies tend to have lower j$_*$ than star-forming galaxies. \citet{Kopenhafer2020ApJ...903..143K} found that breakBRD galaxies tend to be transitioning from star forming to quiescent, even though they currently do not show a broad range of SFRs. Therefore while splitting the breakBRD sample by SFR is not a physically meaningful exercise and will not allow for straightforward comparison with the mass-matched comparison sample, splitting the sample by j$_*$ allows us to split our sample in a way that may give insight into their evolutionary stage.

 In Figure \ref{fig:jstarhist} we show that the j$_*$ distribution of the breakBRD analogue galaxies is quite similar to that of the central comparison sample.  In order to more clearly see CGM differences correlated with j$_*$, we split our galaxy samples by selecting upper and lower quartiles of the j$_*$ distribution of the central comparison sample of 2132 galaxies. This also selects a similar relative fraction of galaxies from the breakBRD analogue sample (20-25\%).  We note that, as discussed above, for the comparison sample j$_*$ is correlated with sSFR, with the mean sSFR of 7.7 $\times$ 10$^{-11}$ yr$^{-1}$ and 1.1 $\times$ 10$^{-10}$ yr$^{-1}$ for low-j$_*$ and high-j$_*$ galaxies, respectively.  However, the breakBRD sSFRs vary little with j$_*$: 3.4 $\times$ 10$^{-11}$ yr$^{-1}$ and 3.1 $\times$ 10$^{-11}$ yr$^{-1}$ for low-j$_*$ and high-j$_*$ galaxies, respectively.  By comparing the CGM in these subsets of our samples, we can tease out what CGM properties are more strongly correlated with sSFR and what are more correlated with j$_*$ (Section \ref{sec:discuss_trends}).

\subsection{(Cold) Circumgalactic Medium Gas Selection}\label{sec:method_coldgas}

In this work we compare the gas properties in the CGM of breakBRD analogues to the larger galaxy population.  Here we define the CGM for each halo as all gas cells that are part of that halo, at least twice the stellar half-mass--radius away from the center of the galaxy, and not bound to any satellite subhalo (i.e., the ``smooth'' component in  \citet{DeFelippis2020ApJ...895...17D}).

Because the CGM consists of gas at a range of temperatures, in this paper we often consider the cold CGM separately from the hot CGM.  Here we define the cold CGM as gas with temperatures below 10$^5$ K.  For the mass range of our sample (M$_*$ between 10$^{10}$ - 10$^{10.5}$ M$_{\odot}$), this is close to the definition of ``cold'' CGM in D20, defined as temperatures below half the virial temperature.  Rather than use a mass-evolving definition for cold gas as in that work, here we choose 10$^5$ K because this temperature includes UV absorption lines that may be observed for comparison with our predictions about the state of the CGM and because including both ``cool'' and ``cold'' gas in our analysis includes a higher mass fraction of the CGM \citep{TUmlinson_2017ARA&A..55..389T}.  

\section{Global CGM Measures}\label{sec:global}

In this section we compare the global CGM properties of central breakBRD galaxies to the larger population of central galaxies with comparable stellar masses.

\begin{figure}
    \centering
    \includegraphics[scale=0.59, trim= 3mm 0mm 0mm 0mm, clip]{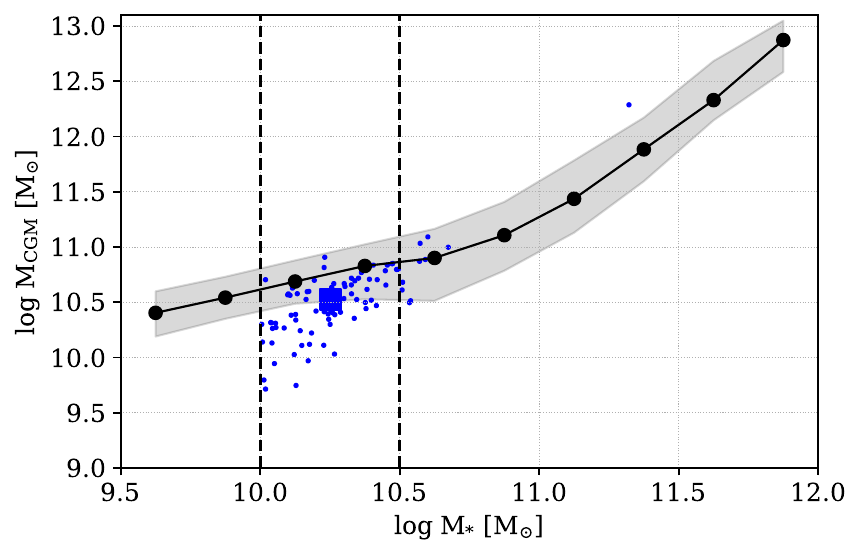}\\
    \includegraphics[scale=0.59, trim= 3mm 0mm 0mm 0mm, clip]{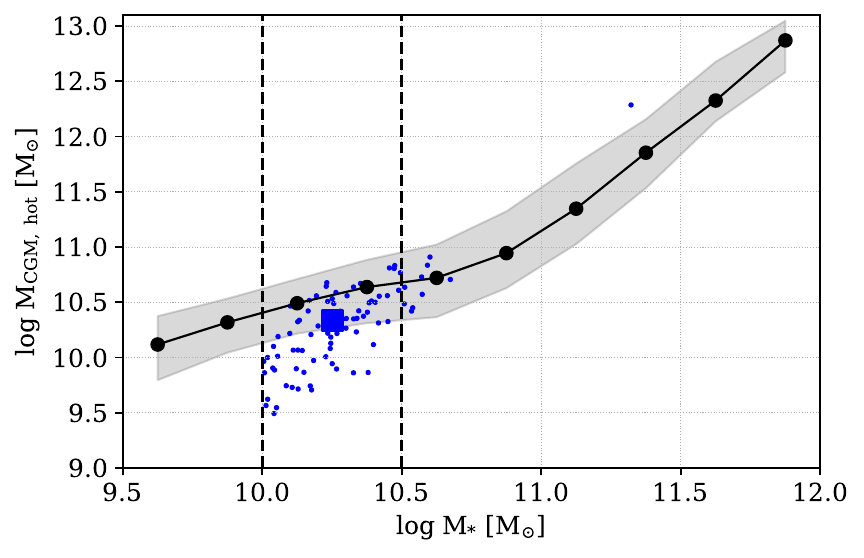}\\
    \includegraphics[scale=0.59, trim= 3mm 0mm 0mm 0mm, clip]{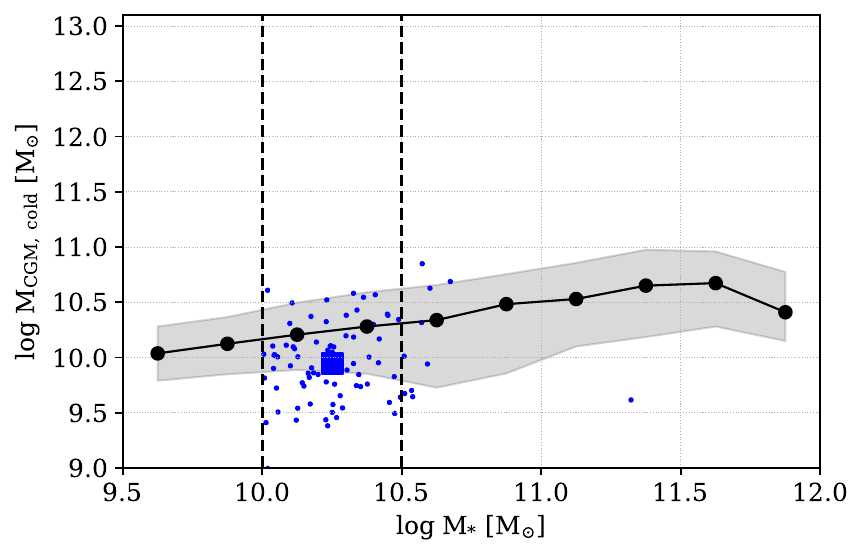}\\

    \caption{The M$_{CGM}$ total (top panel),  M$_{CGM, hot}$ (top panel), and M$_{CGM, cold}$ (bottom panel) vs M$_*$ of central breakBRDs versus the central galaxy sample. The black line and shaded region are the running median and 1-$\sigma$ ranges of the M$_{CGM}$ distribution of all central galaxies and the black points show the center of the stellar mass bins used which are 0.25 dex wide. The breakBRD galaxies are shown as blue points with the median value as the blue square. The comparison sample used in S \ref{sec:maps} is chosen based on stellar mass (10$^{10}$ $\le$ M$_*$/M$_{\odot}$ $\le$ 10$^{10.5}$), and denoted with vertical dashed lines.  Both the hot and cold CGM mass tends to be somewhat lower than the CGM mass of most galaxies at the same stellar mass (KS tests between the breakBRD and comparison galaxies have p-values $<< 0.01$).}
    \label{fig:MCGM_global}
\end{figure}

We first measure the total mass in the CGM, and then split it into cold gas as defined in Section \ref{sec:method_coldgas} and the remaining hot gas. In Figure \ref{fig:MCGM_global} we compare the breakBRD analogues to the central galaxy population.  The breakBRD analogues are shown as the blue points, with the median value as a large blue square.  For comparison, we show the running median in stellar mass bins 0.25 dex wide and 1-$\sigma$ ranges of the distribution of M$_{CGM}$ of the central population as the black line and shaded region, respectively. This allows us to visualize how total stellar mass effects the CGM in the breakBRD mass range while still including a few hundred galaxies in each bin. We also denote the mass range of the comparison sample introduced in Figure \ref{fig:jstarhist} and used in Section \ref{sec:maps} with vertical dashed lines.

From top to bottom we show the total M$_{CGM}$, M$_{CGM, hot}$, and M$_{CGM, cold}$.  In all cases, the breakBRD analogue sample tends to have lower CGM gas masses than the central galaxy sample.  Although the breakBRD analogue sample has a large scatter, their median CGM mass lies at the lower 1-$\sigma$ of the central sample, and the CGM masses are always below the upper 1-$\sigma$ scatter of the central sample (except for the total and hot M$_{CGM}$ of the most massive breakBRD analogue galaxy).  We perform a two-sample Kolmogorov-Smirnov (KS) test to quantify the difference between the breakBRD analogue galaxies and the comparison sample between the dashed lines.  For all three measures of the CGM mass, the $p$-value is orders of magnitude below 0.01, indicating that the CGM mass of breakBRD galaxies and the comparison sample are not drawn from the same distribution.

\begin{figure}
    \centering
    \includegraphics[scale=0.6, trim= 2mm 0mm 0mm 0mm, clip]{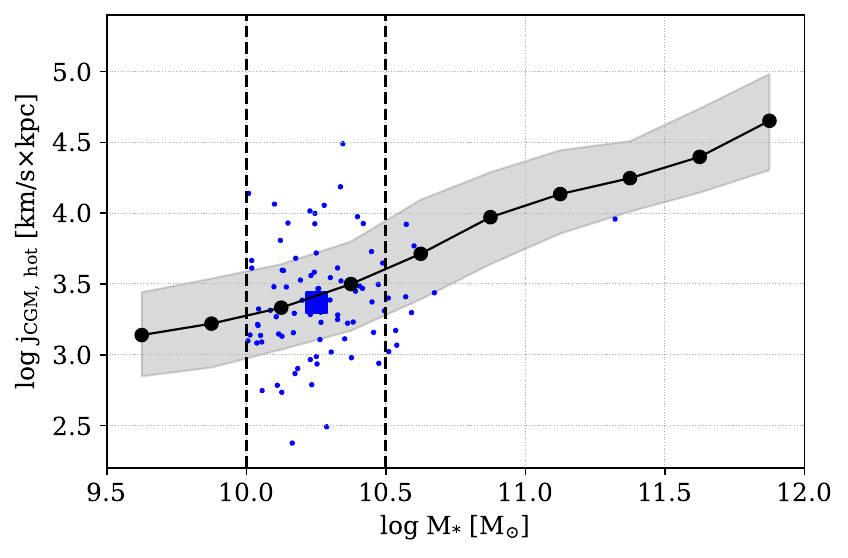}\\
    \includegraphics[scale=0.6, trim= 2mm 0mm 0mm 0mm, clip]{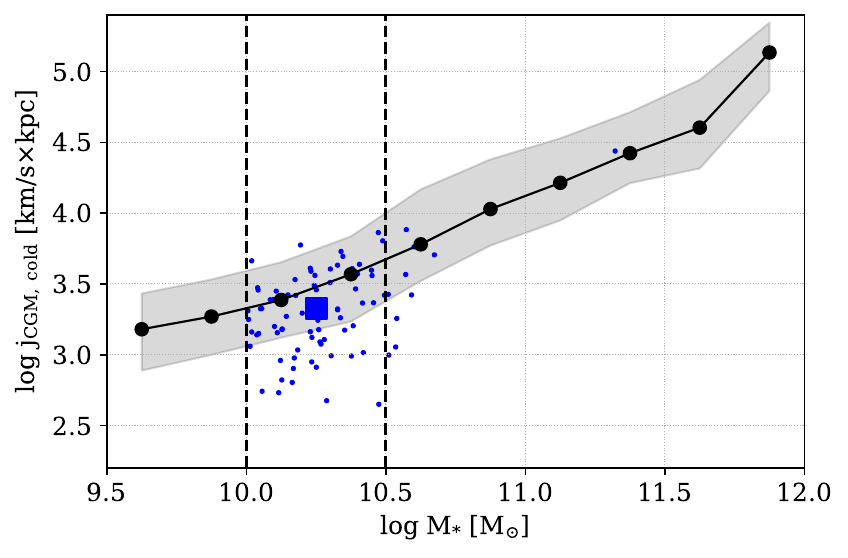}
    \caption{The j$_{CGM, hot}$ (top panel) and j$_{CGM, cold}$ (bottom panel) vs M$_*$ of central breakBRDs versus the central galaxy sample.  The lines and symbols are as in Figure \ref{fig:MCGM_global}. While the j$_{CGM, hot}$ is similar between the breakBRD and comparison samples ($p$-value of 0.4), the j$_{CGM, cold}$ differs ($p$-value well below a threshold of 0.01).}%
    \label{fig:jCGM_global}
\end{figure}

In addition to the gas mass, we compare the angular momentum in the CGM of breakBRD galaxies to the central galaxy samples.  In Figure \ref{fig:jCGM_global} we find that the total angular momentum of hot gas in the CGM is quite similar to the central galaxy sample (top panel), which is quantitatively shown by the KS test $p$-value of 0.4.  However, the angular momentum of cold gas tends to be slightly lower than that of the comparison central galaxy sample, with a $p$-value of 0.0003.  The low j$_{CGM, cold}$ in breakBRD galaxies mainly reflects the low M$_{CGM, cold}$ in the sample.  Indeed, we find that the j$_{CGM}$ - M$_{CGM}$ relation in the breakBRD galaxies is quite similar to the relation in the comparison sample, defined in Section \ref{sec:method_sample} as all central galaxies with M$_*$ between 10$^{10}$ - 10$^{10.5}$ M$_{\odot}$ (not shown).

Figure \ref{fig:jCGM_global} simply sums the total angular momentum of the CGM, and does not account for how the direction of the CGM angular momentum relates to the angular momentum of the stellar disk.  We will discuss angular momentum misalignment in detail for the cold gas in Section \ref{sec:Jmaps}.  

\begin{figure}
    \centering
    \includegraphics[scale=0.595, trim= 2mm 0mm 2mm 0mm, clip]{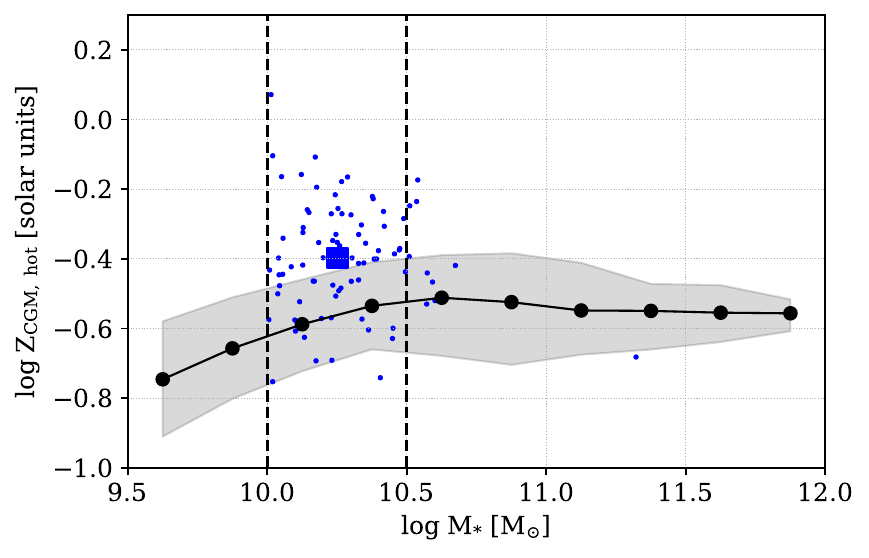}\\
    \includegraphics[scale=0.595, trim= 2mm 0mm 2mm 0mm, clip]{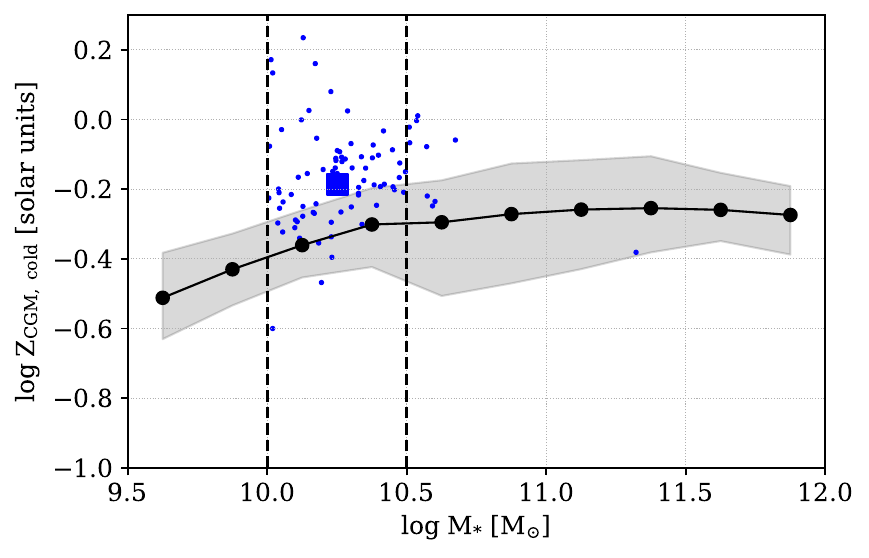}
    \caption{The metallicity of hot (top panel) and cold (bottom panel) CGM gas vs M$_*$ of central breakBRDs versus the central galaxy sample. A KS test comparing the breakBRD analoque sample to the comparison sample results in $p$-values well below a 0.01 threshold.}%
    \label{fig:ZCGM_global}
\end{figure}

\begin{figure*}
    \centering

    \includegraphics[scale=0.5, trim= 0mm 0mm 0mm 0mm, clip]{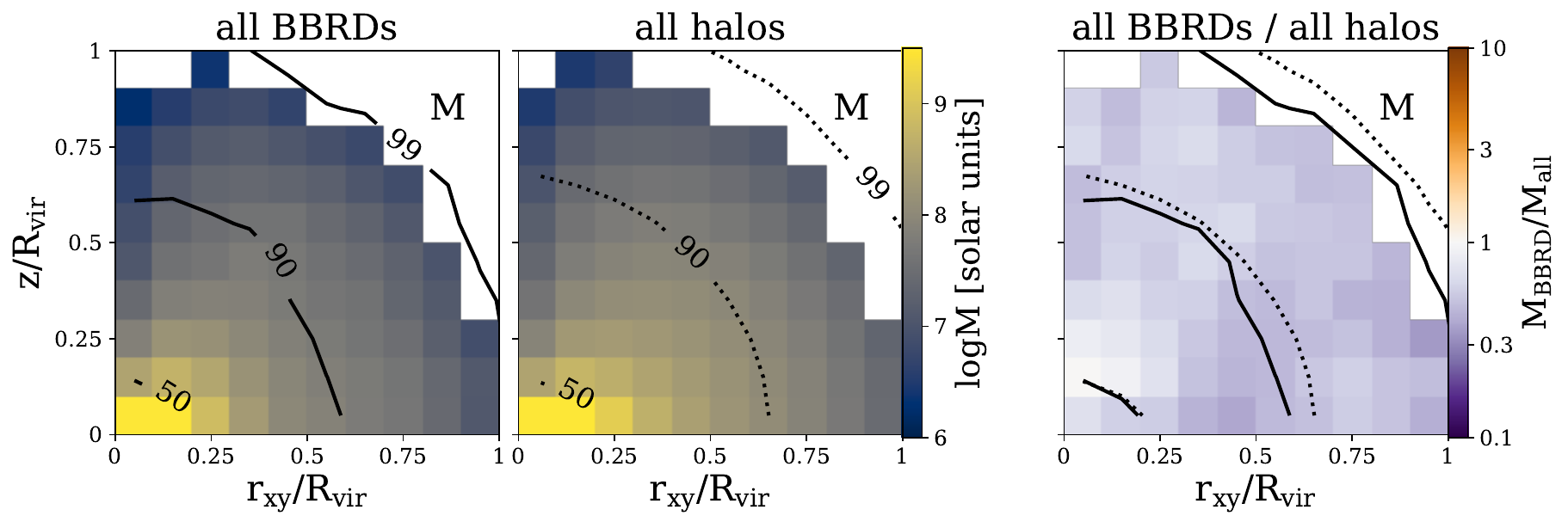}\\    
    \includegraphics[scale=0.5, trim= 0mm 0mm 0mm 0mm, clip]{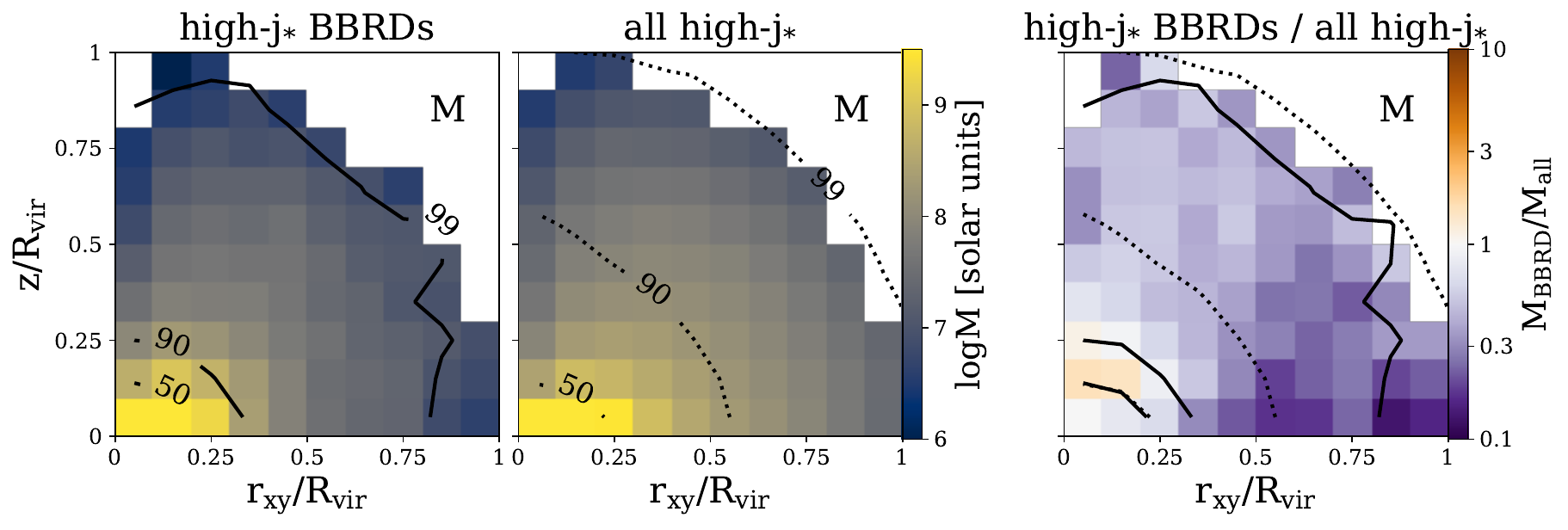}\\
    \includegraphics[scale=0.5, trim= 0mm 0mm 0mm 0mm, clip]{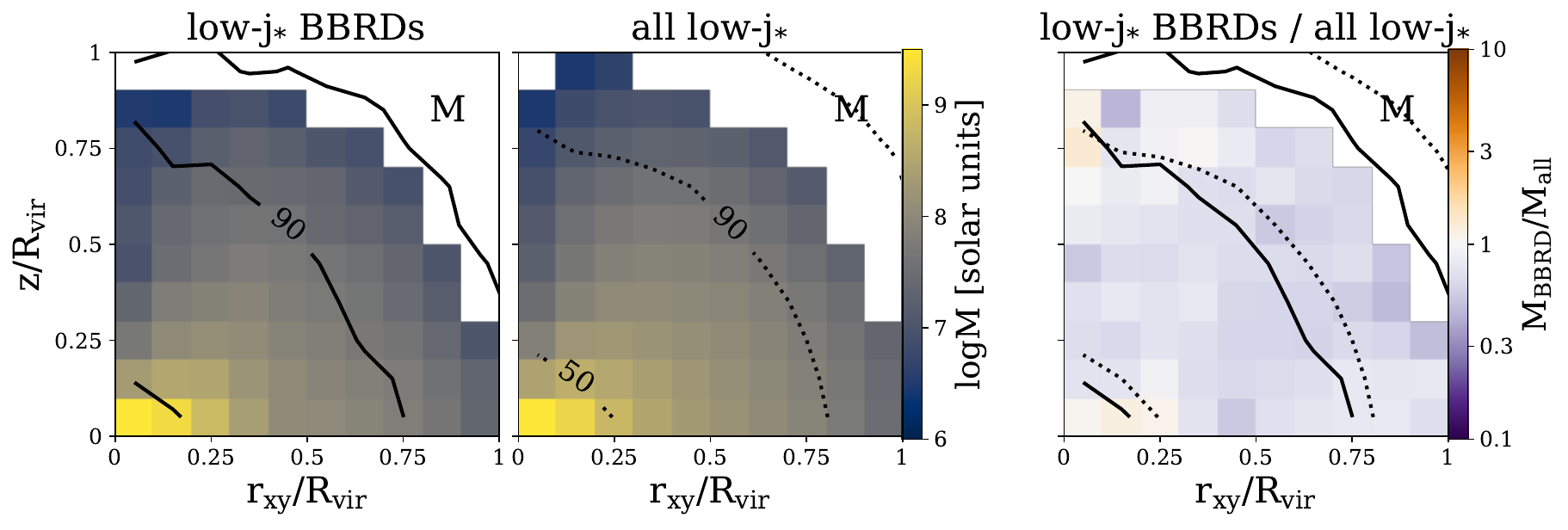}\\ 
    
    \caption{The distribution of cold gas mass in breakBRD analogue versus central galaxies.  From left to right the panels show breakBRD galaxies, the comparison sample (central galaxies with stellar masses between 10$^{10}$ - 10$^{10.5}$ M$_{\odot}$), and the ratio of the mass distribution.  The upper panels show all BBRD galaxies (except the one with a stellar mass above 10$^{11}$ M$_{\odot}$) and all galaxies in the comparison sample. The middle panels only show those galaxies in the high-j$_*$ samples, and the bottom panels only show galaxies in the low-j$_*$ sample. The black lines in each panel are isodensity contours of cold gas that are labeled by the percentage of cold gas mass they enclose. They are the same in all subsequent figures. The CGM of breakBRD galaxies tends to have less mass throughout, and be slightly more concentrated than the comparison sample. 
    }
    
    \label{fig:MCGM_mapall}
\end{figure*} 

Finally, we examine the metallicity of the CGM gas in breakBRD galaxies.  In Figure \ref{fig:ZCGM_global} we show the mass-weighted average metallicity in the hot and cold CGM.  The metallicity of both the hot and cold gas in the CGM of breakBRD galaxies tends to be higher than in the total central sample.  In fact, the difference in the CGM metallicity between breakBRD and central galaxies is more significant than either the CGM mass or the j$_{CGM}$, with the average breakBRD value higher than the 1-$\sigma$ region of the central sample. This is also reflected by the smallest $p$-values in a two-sample KS test, again several orders of magnitude below a 0.01 threshold indicating different distributions.

\begin{figure*}
    \centering
    \includegraphics[scale=0.5, trim= 0mm 0mm 0mm 0mm, clip]{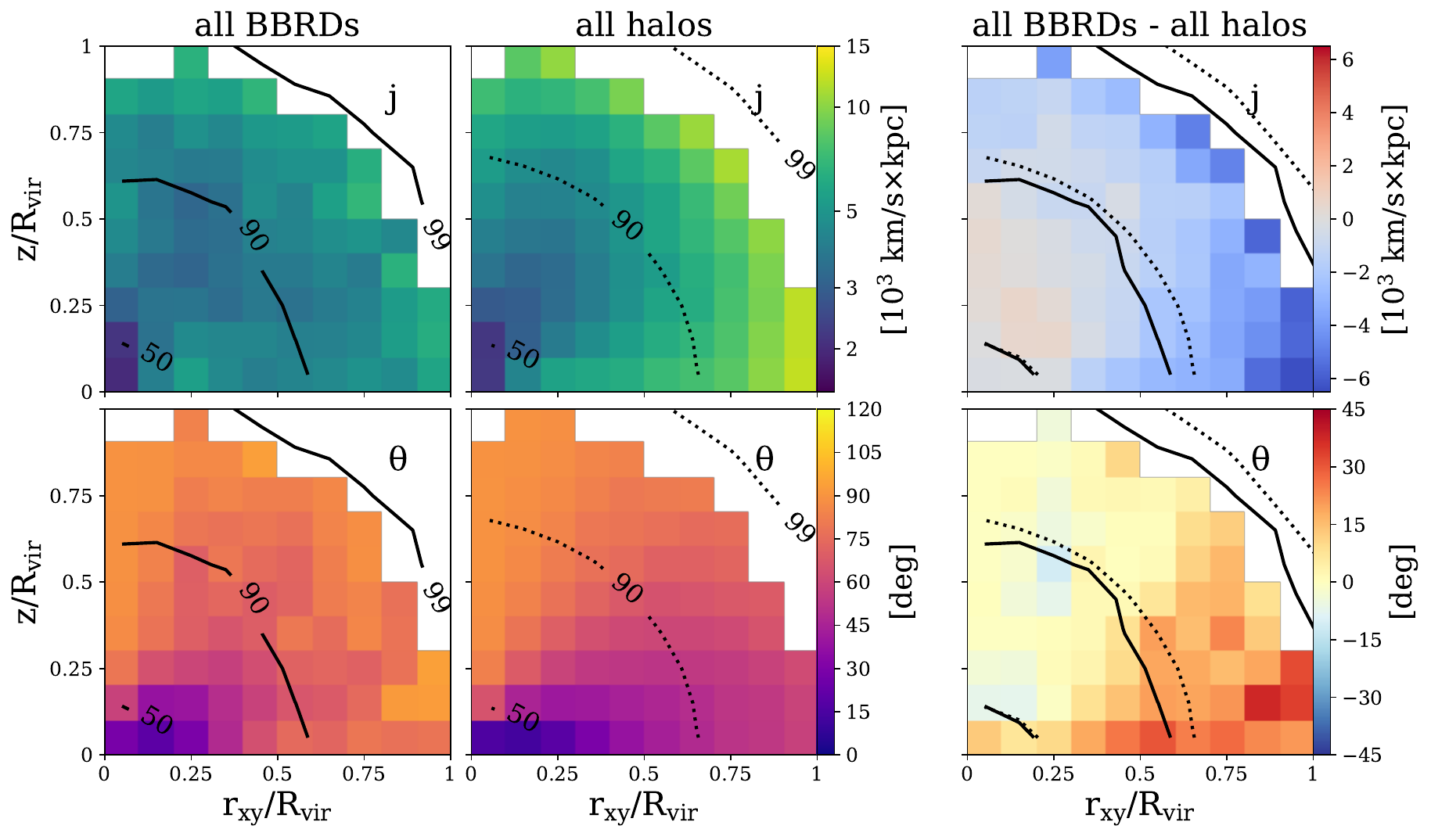}\\    

    \caption{\textbf{Top panels:}  The angular momentum distribution of the cold gas in the CGM of breakBRD versus comparison galaxies.  \textbf{Bottom panels:} The misalignment angle of j$_*$ and j$_{CGM}$ of the cold gas in the CGM of breakBRD versus central galaxies.  Angular momentum is low and less well aligned with the stellar component in breakBRD galaxies.
    }
    \label{fig:jCGM_mapall}
\end{figure*}

\begin{figure*}
    \centering

    \includegraphics[scale=0.5, trim= 0mm 0mm 0mm 0mm, clip]{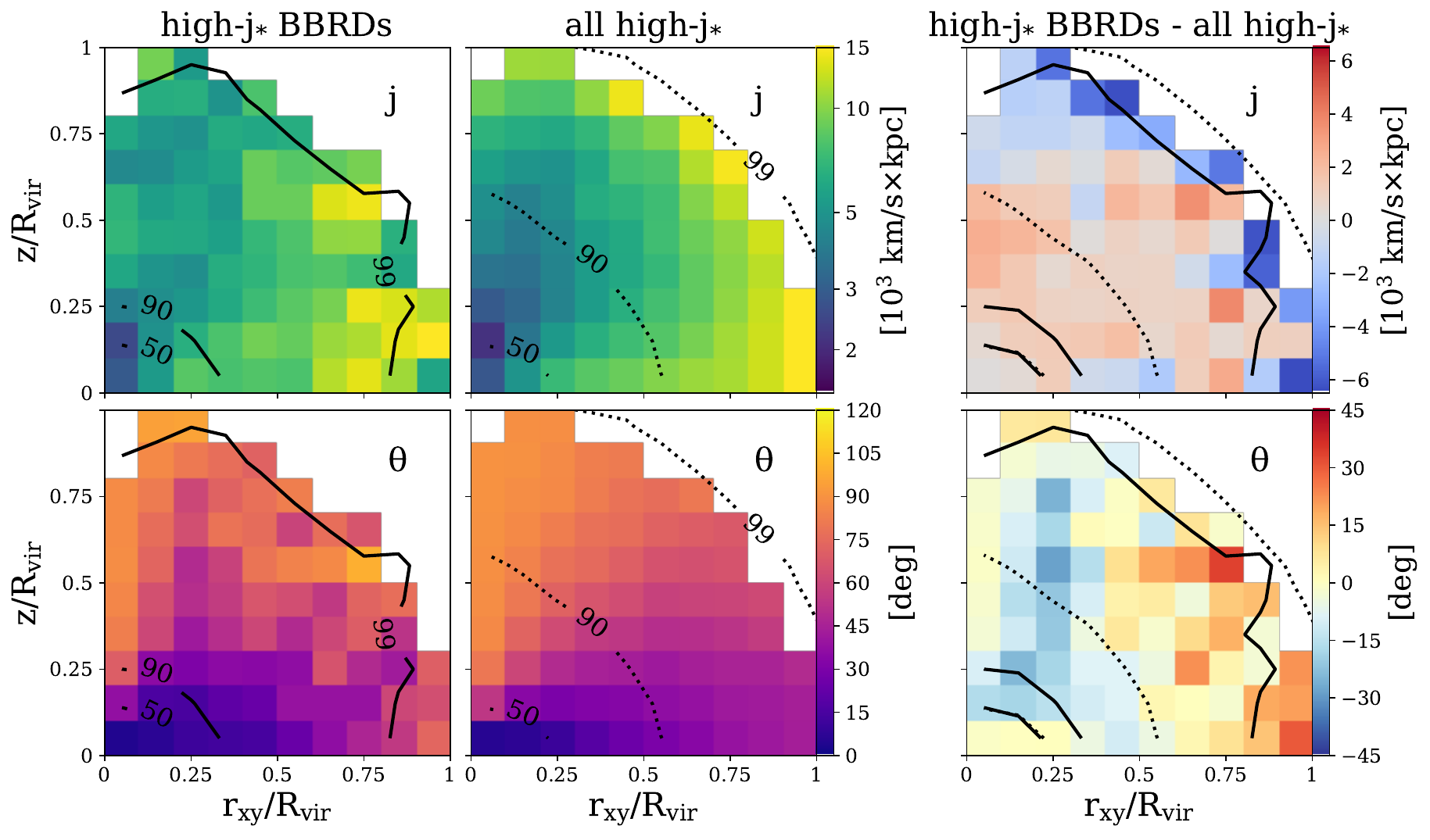}\\
    \includegraphics[scale=0.5, trim= 0mm 0mm 0mm 0mm, clip]{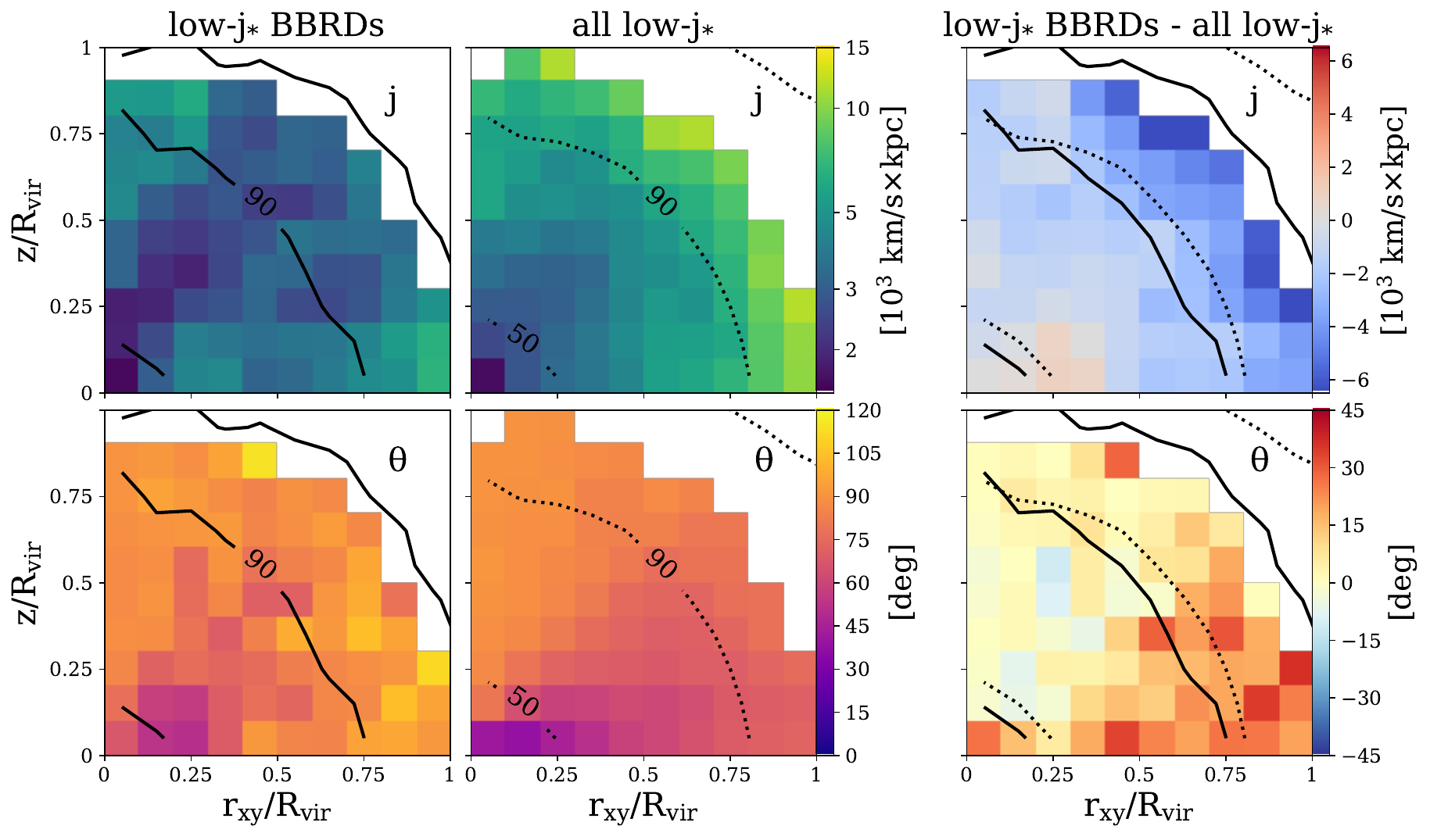}\\    

    \caption{The angular momentum distribution and alignment angle (j$_{CGM}$ versus j$_*$) of cold gas in breakBRD versus comparison galaxies, split into high-j$_*$ (top panels) and low-j$_*$ quartiles (bottom panels).  The low-j$_*$ breakBRDs show lower angular momentum and more misalignment than the comparison sample, reflecting the breakBRD population as a whole (Fig \ref{fig:jCGM_mapall}).}
    \label{fig:jCGM_mapjsplit}
\end{figure*}

\section{The Distribution of the CGM surrounding BreakBRDs}\label{sec:maps}

In this section we examine the distribution of mass, angular momentum, and metals in the CGM in detail. We assume that the CGM around galaxies is symmetric above and below the disk as well as azimuthally, and varies as a function of cylindrical radius and height above the disk (in Appendix \ref{app:CGMsymmetry} we show the average variation from symmetry for our samples). We then map the CGM properties onto a grid with [$z/$R$_{vir}$, $r_{xy}/R_{vir}$] cells that are [0.1,0.1] on each side. Cells are only shaded if they are $\leq$ R$_{vir}$ from the disk and at least 25\% of galaxies in the sample have mass in that region of their CGM. In addition to showing projected maps, we also show either the difference or ratio between the breakBRD and comparison sample in the rightmost plot in each row. 

As we define in Section \ref{sec:method_sample}, the comparison sample includes all central galaxies with stellar masses between 10$^{10}$ - 10$^{10.5}$ M$_{\odot}$.  Here we also split the breakBRD and comparison sample into the high-j$_*$ and low-j$_*$ subsamples that are determined by the upper and lower j$_*$ quartiles of the comparison sample.  As discussed in Section \ref{sec:method_sample}, because the breakBRD sample has a similar j$_*$ distribution to the comparison sample, these j$_*$ values each select 20-25\% of the breakBRD sample as well.

\subsection{Mapping the Mass distribution}

We first compare the mass distribution of the CGM of the breakBRD and comparison central galaxy samples.  In Figure \ref{fig:MCGM_mapall} we plot the average mass distribution of the cold gas in the CGM of galaxies in the breakBRD sample and the central galaxies. The black lines are isodensity contours enclosing different percentages of the cold gas mass, which aid comparison of the CGM density profiles of different samples.

When we first look at breakBRD galaxies as a whole compared to the entire comparison central galaxy sample (top panels), we see that on average there is less cold gas mass throughout the halo.  We also see some indication that the cold gas is less extended in breakBRD galaxies than the comparison sample by comparing the solid (BBRD galaxies) and dotted (comparison galaxies) lines denoting 90\% and 99\% of the CGM mass.  
The somewhat lower cold CGM mass throughout the halo agrees with Figure \ref{fig:MCGM_global}, which finds lower M$_{CGM, cold}$ in breakBRD galaxies relative to the comparison sample. In Appendix \ref{app:variationbtwngals} we show the variation at any grid cell in our maps is comparable to the cold gas mass itself, indicating that the difference in the populations could be overlooked in comparisons between individual galaxies.

When we split the sample into the high- and low-j$_*$ samples we find that the high-j$_*$ breakBRD galaxies have more centrally concentrated mass than comparison high-j$_*$ galaxies, as seen most clearly by comparing the 90\% mass lines.  Although beyond 0.25 R$_{vir}$ there is universally less mass in the CGM of breakBRD analogues, a somewhat bigger difference can be seen beyond 0.5 R$_{vir}$ near the disk plane.
The low-j$_*$ mass distribution is very similar in the two populations, although again breakBRD analogue galaxies seem to have a slightly more centrally concentrated CGM and less overall cold CGM mass.

\subsection{Mapping the CGM Angular Momentum}\label{sec:Jmaps}

As with the mass distribution, we map the specific angular momentum of CGM gas in Figures \ref{fig:jCGM_mapall} \& \ref{fig:jCGM_mapjsplit} (and discuss variation within the populations in Appendix \ref{app:variationbtwngals}).  Because the angular momentum can be in any direction, we also show the angular offset between the CGM angular momentum vector and the stellar angular momentum vector in the lower panels.  

In Figure \ref{fig:jCGM_mapall} we find that breakBRD analogue galaxies tend to have lower specific angular momentum values than the comparison sample, particularly beyond 0.5 R$_{vir}$.  The relative angular momentum in the CGM of breakBRD galaxies shows the most difference from the comparison sample farther out in the halo where there is much less mass, and near the disk plane.  In the comparison sample, j$_{CGM}$ increases near the disk plane, unlike the breakBRD galaxies' j$_{CGM}$.  In addition to a lower magnitude, the direction of j$_{CGM}$ in breakBRD galaxies is generally more misaligned than the comparison sample.  The misalignment difference is the most dramatic near the disk plane, and increases along the plane towards larger cylindrical radius.  

In Figure \ref{fig:jCGM_mapjsplit} we show the angular momentum maps for the high-j$_*$ and low-j$_*$ galaxy samples.  In the high-j$_*$ galaxies, there are no strong differences, or consistent spatial trends when comparing the two samples, although there is a hint that the j$_{CGM}$ is slightly larger in the breakBRD galaxies.  We highlight that the j$_{CGM}$ and angular momentum distribution in the high-j$_*$ breakBRDs show less smooth gradients than the comparison sample, but that may be somewhat due to the much smaller sample size. 

On the other hand, low-j$_*$ breakBRD galaxies show systematically different j$_{CGM}$ maps than the low-j$_*$ comparison sample.  First, the j$_{CGM}$ magnitude is smaller in the breakBRD galaxies, with the difference increasing with increasing radius.  Second, low-j$_*$ breakBRD galaxies show a much stronger misalignment between their j$_{CGM}$ and j$_*$ vectors than the comparison sample, with some cells even showing counter-rotation.  Indeed, the comparison between the low-j$_*$ galaxies in the rightmost panels looks similar to the comparison between the total samples in Figure \ref{fig:jCGM_mapall}. This hints that the higher angular momentum and alignment in high-j$_*$ breakBRD galaxies may be unusual for the population.

\subsection{Mapping the CGM Metallicity}

\begin{figure*}
    \centering
    \includegraphics[scale=0.5, trim= 0mm 0mm 0mm 0mm, clip]{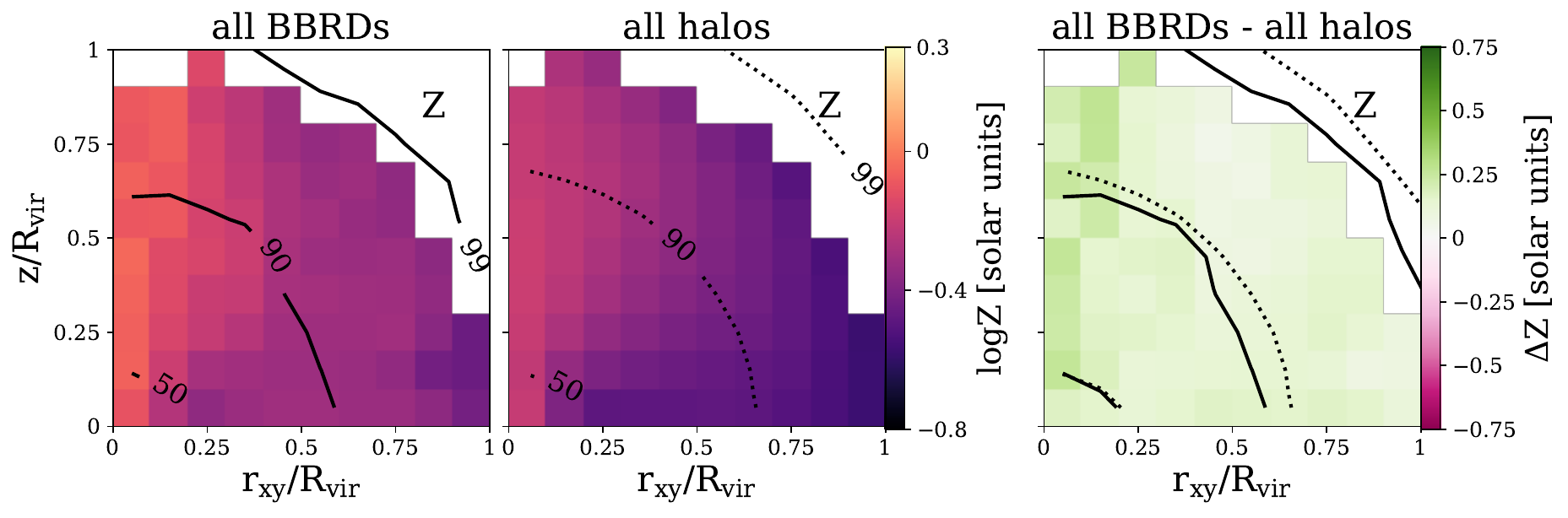}\\    
    \includegraphics[scale=0.5, trim= 0mm 0mm 0mm 0mm, clip]{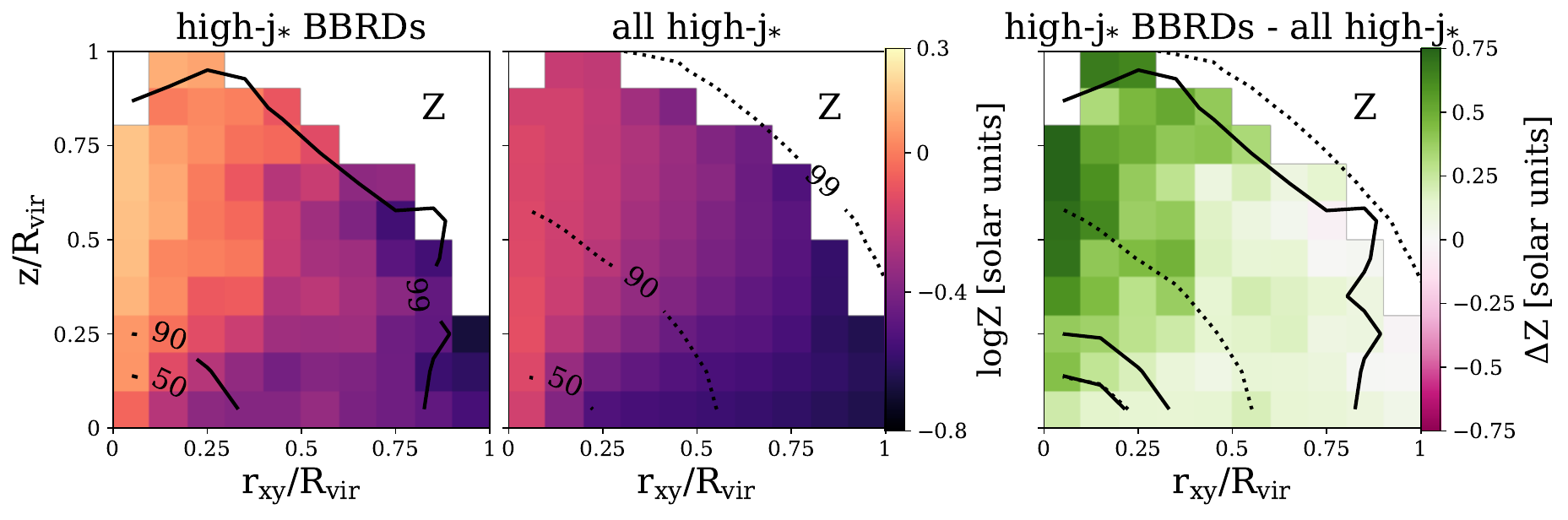}\\
    \includegraphics[scale=0.5, trim= 0mm 0mm 0mm 0mm, clip]{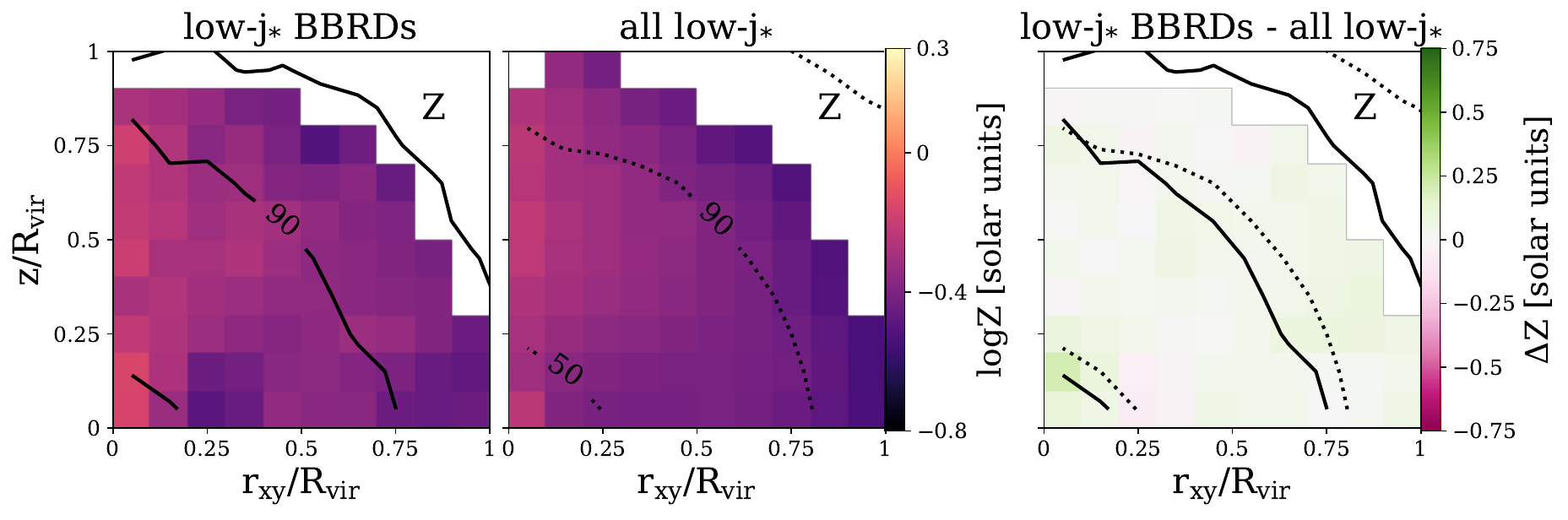}\\    

    \caption{The metallicity distribution of cold gas in breakBRD versus comparison galaxies, with the complete samples in the top panels, and high-j$_*$ and low-j$_*$ samples in the middle and bottom panels, respectively.  The metallicity is clearly the highest in high-j$_*$ breakBRD galaxies along the minor axis.}
    \label{fig:ZCGM_mapall}
\end{figure*}

Finally, in Figure \ref{fig:ZCGM_mapall} we show the metallicity distribution in the CGM of breakBRD galaxies and the comparison sample.  As above, we first focus on the total populations before splitting them into high- and low-j${_*}$ samples.  

In all galaxies, we see that higher metallicity gas lies along the pole and lower metallicity gas lies closer to the disk plane, in agreement with previous work \citep[e.g.][]{Truong_2021MNRAS.508.1563T}.  However, in breakBRD galaxies, the CGM metallicity is consistently similar to or higher than in the comparison sample, as evidenced by the top right panel (albeit with galaxy-to-galaxy scatter, as shown in Appendix \ref{app:variationbtwngals}).  Unlike the mass and angular momentum distributions, there is no radial dependence along the disk plane on the metallicity difference between breakBRD and comparison galaxies.  

The high-j$_*$ sample shows similar results, with a pronounced difference in the CGM metallicity of breakBRD and comparison galaxies, particularly along the polar axis and out to about 30$^\circ$ from the minor axis. Low-j$_*$ galaxies show similar metallicities in the breakBRD and comparison samples, showing no trend radially or as we look towards the polar or disk axes.

\section{Discussion}\label{sec:discussion}

In this section we first attempt to connect the breakBRD CGM properties to the star formation and gas distribution in breakBRD disks, then use the differences in the SFRs and j$_*$ of the different breakBRD and comparison subpopulations to discuss what galaxy properties are most closely tied to CGM mass, angular momentum and metallicity.  Finally, we make some general predictions for future observations of the CGM around central breakBRD galaxies.

\subsection{The CGM as part of the breakBRD gas cycle}\label{sec:discussion_gascycle}

In \citet{Kopenhafer2020ApJ...903..143K}, we found that breakBRD analogue galaxies have a centrally-concentrated star formation distribution.  This comes as no surprise as they are defined to have red disks and star formation in the central 2 kpc.  Importantly, we found that gas is also concentrated within the central 2 kpc.  This concentration is due to low gas content beyond 2 kpc and normal gas mass in the central 2 kpc.  We found similar results when looking specifically at the dense gas (star forming gas in TNG).  In addition, breakBRD galaxies identified at $z=$ 0 and tracked backwards have tended to lose gas mass since $z=$ 0.5, particularly in their outskirts (beyond 2 R$_{1/2}$), while other galaxies have tended to gain gas mass.  

While the CGM is an important reservoir of gas moving into and out of the disk, the timescales connecting gas in the outer CGM to the disk ISM can be long \citep[e.g.][]{Christensen_2016ApJ...824...57C}.  Therefore, in this section we pose connections between the CGM and disk gas distributions, but specifically tracking gas flows through the CGM into or out of the disk is beyond the scope of this paper.  Despite this caveat, we can relate the differences between the CGM in the breakBRD and comparison sample in this paper to the galaxy properties found in  \citet{Kopenhafer2020ApJ...903..143K}, and attempt to gain insight into the gas cycle of breakBRD galaxies. 

First, the high CGM metallicity in breakBRD galaxies is likely to indicate some combination of three scenarios: in the past more metals were being added to the CGM, less metals were (and are) leaving the CGM, or less low-metallicity gas is accreting into the system.  Given that the mass in the CGM is low, trapping an excess of metal-rich gas is disfavored.  Because the global SFR in breakBRD galaxies is low (K20), we do not think that an excess of metals are being added to the CGM.  The enhanced metallicity along the polar axis, particularly of the high-j$_*$ sample, could be from a small amount of metals added to the already low mass in the CGM.  Therefore, we tentatively contend that the most likely scenario is that less low-metallicity gas historically accreted into the CGM of breakBRD galaxies, which may have resulted in higher metallicities even in the star-forming and ejected gas.  This scenario would also likely be reflected in an enhanced metallicity in the surviving disk, so could be tested in future work.

The low angular momentum in the gas may also be related to a lower accretion rate into the CGM of breakBRD galaxies.  The lower angular momentum in the total sample is most clearly seen in the CGM outskirts and near the disk plane, where we might expect accretion from the IGM to deposit mass and momentum.  Interestingly, the high-j$_*$ galaxies actually show higher angular momentum that is generally aligned with j$_*$.  This may highlight the important role that feedback can play in maintaining angular momentum in the CGM, although we expect that much of the angular momentum in the CGM is gained from IGM accretion \citep{DeFelippis_2017ApJ...841...16D}.

We now briefly look towards the future evolution of breakBRD analogues.  The low CGM mass, particularly low cold gas mass, may indicate that these galaxies will run out of fuel for star formation and quench.  This would agree with the \citet{Davies_EAGLECGM_2020MNRAS.491.4462D, Davies_2019MNRAS.485.3783D} results that central galaxies with low CGM gas fractions in TNG (and EAGLE) are more likely to be quenched.  Indeed, K20 found that the majority of $z=$0.5 breakBRD analogues are quenched by $z=$0.  Interestingly, the breakBRD sample shows a lower angular momentum than the comparison sample, while \citet{Lu_TNGAM_2022MNRAS.509.2707L} find that low SFR and quenched galaxies tend to have higher CGM angular momentum.  We note that in that work the difference in the samples becomes clear at M$_*$ $>$ 10$^{10.5}$ M$_{\odot}$.  

In addition to the mass in the CGM being somewhat more centrally concentrated, the combination of lower angular momentum and higher misalignment between j$_{CGM}$ and j$_*$ indicates that gas is more likely to enter the disk in the central regions \citep{Trapp_FIREAM_2022MNRAS.509.4149T}.  Because the timeframe for infall from the CGM to the disk could be long \citep{Oppenheimer_2010MNRAS.406.2325O,Ford_2014MNRAS.444.1260F,Christensen_2016ApJ...824...57C}, and the low mass and angular momentum misalignment extend to the CGM outskirts, we would expect the centrally-star forming phase to be long-lived. Recall that K20 found that $\sim$86\% of central breakBRD galaxies identified at $z = 0.5$ quenched by $z = 0$ (compared to $\sim$25\% of the parent central sample), while $\sim$26\% of central breakBRD analogue galaxies identified at $z = 0.1$ have quenched by $z = 0$ (compared to $\sim$4.5\% of the parent central sample).  Therefore we predict that these galaxies will remain centrally star-forming until they become passive.

\subsection{Connecting the CGM to Galaxy Properties}\label{sec:discuss_trends}

As we discussed in Section \ref{sec:method_sample}, the central breakBRD sample has a similar distribution of j$_*$ to the comparison sample, but a lower average sSFR than even the low-j$_*$ comparison sample.  By comparing the different galaxy subsets in Section \ref{sec:maps} we can briefly discuss what aspects of the CGM seem more unique to breakBRD galaxies, and what aspects are more closely connected to the j$_*$ or sSFR of galaxies.  

First, we note that the CGM mass around breakBRD galaxies is lower than either high- or low-j$_*$ comparison galaxies.  As breakBRD galaxies have a lower median sSFR than either of the comparison subsamples, this agrees with other work that galaxies with lower SFRs have low CGM masses \citep{Davies_EAGLECGM_2020MNRAS.491.4462D, Davies_2019MNRAS.485.3783D}.  Therefore, we argue that M$_{CGM}$ is likely more dependent on sSFR than on j$_*$.  

On the other hand, there are identifiable differences in the j$_{CGM}$ around breakBRD galaxies with high-j$_*$ versus low-j$_*$.  Reflecting the larger comparison sample, breakBRD high-j$_*$ galaxies have higher j$_{CGM}$ that is better aligned with j$_*$.  As one might expect, j$_{CGM}$ seems more strongly correlated with j$_*$ than with the sSFR of galaxies.  

The metallicity is where our intuition and previous work does not align with our results.  Previous work studying simulated galaxies in TNG has found that galaxies with higher sSFR tend to have stronger azimuthal metallicity gradients, with high metallicity along the minor axis and low metallicity near the galaxy plane \citep{Truong_2021MNRAS.508.1563T}.  Indeed, that is what we see in the comparison sample, using our knowledge that the high-j$_*$ sample has higher sSFR than the low-j$_*$ sample.  However, surprisingly, the highest metallicity along the minor axis and highest gradient from minor to major axis is in the high-j$_*$ breakBRDs, which have lower sSFR than either of the comparison subsamples.  Not only this, but low-j$_*$ breakBRDs, which have similar sSFRs to high-j$_*$ breakBRDs, do not show a metallicity gradient.    

This may indicate that the metallicity gradient is actually more dependent on j$_*$ than sSFR.  However, because the j$_*$ distribution is quite similar between breakBRDs and the comparison sample, perhaps the metallicity gradient is most closely tied to j$_{CGM}$.  In the future this could be tested in TNG using larger samples with low sSFR and high j$_{CGM}$.  This would tell us whether or not the strong metallicity gradient could be unique to breakBRD galaxies, and therefore perhaps closely connected to the gas and SF concentration in the disk.

\subsection{Observational Predictions}\label{sec:discussion_obs}

While we do not make mock observations of our simulations, in this section we attempt to synthesize our results into observational predictions for the CGM around central breakBRD galaxies.  

First, we expect observations to find a more rapid fall-off of CGM absorbers in number and strength as a function of impact parameter around breakBRD galaxies.  This is due to the less extended cold gas mass in the CGM, seen in all central BBRDs, but most dramatically in high-j$_*$ BBRDs.  We note that as observations of the CGM in emission are likely only able to map the CGM closer to galaxies we would not expect a discernible difference based on gas mass.  

Based on the angular momentum of the cold CGM we predict that the velocities of absorbers in breakBRD galaxies will be lower than average.  However, we see that the angular momentum is actually somewhat higher for high-j$_*$ BBRD galaxies, showing that the scatter is large.

We find that for most breakBRDs, the CGM will not be corotating with the disk, and in some cases, particularly in low-j$_*$ galaxies, regions of the CGM are counter-rotating with respect to the stellar disk. This differs both from the other galaxies in TNG100 and from observations:  we predict observations will not find a thick corotating “disk” of halo gas aligned with the galaxy disk \citep{Steidel_2002ApJ...570..526S,Ho_2017ApJ...835..267H,DiamondStanic_2016ApJ...824...24D}.

Although we find that breakBRD analogues have a more metal-rich CGM than the comparison sample, there is still a significant overlap in CGM metallicity. Because of this, we expect that the difference in CGM metallicities would require either a large sample of absorption sight-lines or emission mapping of the CGM.  However, we do predict that there is likely to be a more significant difference along the minor axis where the most metal-rich gas is found in all galaxies.  

We note that these observational predictions must be for a population comparison - as shown in the distribution of global values in Figures \ref{fig:MCGM_global} - \ref{fig:ZCGM_global} and in the maps of the variations between galaxies shown in Appendix \ref{app:variationbtwngals}, there is overlap between the CGM properties of breakBRD galaxies and the larger population.

\section{Conclusion}\label{sec:conclusions}

In this paper we have studied the CGM of central breakBRD analogue galaxies identified in \citet{Kopenhafer2020ApJ...903..143K}.  We have found a number of differences in the CGM gas properties of breakBRD galaxies from the general sample, even when splitting the populations into high-j$_*$ and low-j$_*$ subsamples.  Our main results are listed below:\\
\begin{enumerate}
    \item BreakBRD galaxies tend to have a lower mass CGM, particularly when only considering cold CGM gas (Figure \ref{fig:MCGM_global}).  The cold gas in the CGM of breakBRDs is on average less extended than in a comparison sample with similar stellar mass (Figure \ref{fig:MCGM_mapall}).
    \item{The angular momentum in the CGM of breakBRD galaxies is slightly low, mainly in cold gas (T $<$ 10$^5$ K) (Figure \ref{fig:jCGM_global}). }
    \item{By mapping the CGM angular momentum, we find that the most dramatic differences between the breakBRD and comparison sample are located near the galaxy disk, both in terms of lower angular momentum, and more misalignment between j$_{CGM}$ and j$_*$ (Figure \ref{fig:jCGM_mapall}).  We also find stronger systematic differences between the CGM of low-j$_*$ breakBRDs and comparison galaxies than in the CGM of high-j$_*$ galaxies (Figure \ref{fig:jCGM_mapjsplit}).}
    \item{The metallicity of the CGM in breakBRDs is higher than in central TNG galaxies (Figure \ref{fig:ZCGM_global}).  This higher metallicity is seen throughout the CGM in BBRDs, with the most dramatic increase seen near the poles in high-j$_*$ BBRDs (Figure \ref{fig:ZCGM_mapall}).}
\end{enumerate}
 
Together these differences between the CGM in breakBRD and comparison galaxies indicate that the lack of gas and SF in the outskirts of breakBRD galaxies could be connected to the state of the CGM:  there is not enough gas in the CGM to sustain high SFRs through infall, yet the angular momentum is low and/or misaligned so gas reaching the disk is likely to do so near the center of the galaxy.  The high metallicity in the CGM of breakBRD galaxies could be achieved by a small amount of metals mixing into the low-mass CGM.  Overall, we argue that most of the unique properties of both the disk and CGM of breakBRD galaxies could be traced to low IGM accretion into the CGM (Section \ref{sec:discussion_gascycle}).

We note that not all of our results are clearly consistent with this scenario, however.  In particular we highlight the strong azimuthal metallicity gradient seen in high-j$_*$ breakBRD analogues that is missing in the more strongly star-forming comparison samples as well as in the low-j$_*$ breakBRD sample with similar SFRs.  In Section \ref{sec:discuss_trends} we discuss whether this may be connected to the higher j$_{CGM}$ in high-j$_*$ breakBRDs.
 
Opportunely, the high metallicity of the breakBRD CGM may allow for these differences to be detectable in the near future. Taking advantage of current measurements in absorption that allow line of sight detections, a proposed cubesat called Maratus would map in emission the extended CGM around past pencil beam detections in the far ultraviolet (Tuttle et al, in prep). Using primarily OVI, expected to be the brightest tracer of $10^5-10^6$ K gas, the hope is that this proof of principal instrument would pave the way for upcoming large scale missions. Although directly measuring the angular momentum component would be complex, the differences in CGM metallicity as a function of galaxy properties will be important for disambiguating local and large scale trends.

\acknowledgments
The authors first thank the referee, whose comments improved the paper.  The authors would also like to thank the
IllustrisTNG collaboration for making their data public.   The data used in this work were hosted on facilities supported by the Scientific
Computing Core at the Flatiron Institute, a division of the
Simons Foundation, and the analysis was largely done using
those facilities. SET acknowledges support from the National Science Foundation through grant AST-1813462. DD acknowledges support from the ANR 3DGasFlows Project (ANR-17-CE31-0017).

\bibliography{breakBRD_bib}{}
\bibliographystyle{aasjournal}

\begin{appendix}
\restartappendixnumbering

\section{Variation between Galaxies}\label{app:variationbtwngals}

The CGM distribution maps in Figures \ref{fig:MCGM_mapall} - \ref{fig:ZCGM_mapall} are averages over a population of galaxies.  These galaxy populations must also have variations in their CGM distributions, as evidenced by the range of global properties in Figures \ref{fig:MCGM_global} - \ref{fig:ZCGM_global}.  In this Appendix we show the variations in the CGM maps of the breakBRD and comparison populations.  For each CGM property: mass, angular momentum, and metallicity, we take the average map of every galaxy in each sample and then find the 1-$\sigma$ dispersion between the galaxies.  

\begin{figure}
    \includegraphics[scale=0.41]{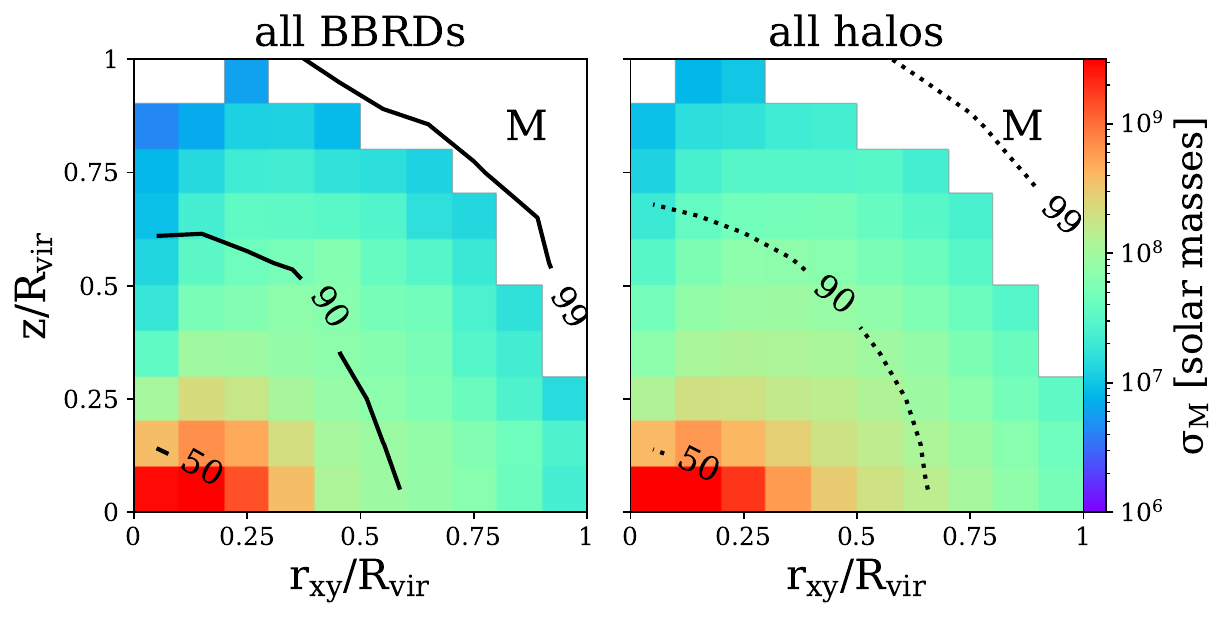}
    \caption{The variation of the cold gas mass distribution in the CGM between galaxies in both the breakBRD (left) and comparison (right) populations in each mapped cell.   } \label{fig:app_massbetweengals}
\end{figure}

In Figure \ref{fig:app_massbetweengals} we map the 1-$\sigma$ variation of the cold gas mass in each of grid cells described in Section \ref{sec:maps}.  First, we note that both the breakBRD and comparison samples show quite similar variations in the cold gas mass distribution.   Although the $\sigma_{\rm M}$ have the highest magnitude in the central regions, the fractional variation is lowest near the galaxy center, and increases to the outskirts with a maximum variation of about a factor of 2.  This is expected due to the steeply declining CGM mass as a function of radius - a small difference in slope will lead to large differences in the cold gas mass at larger radii.  We also see the same behavior in the high-j$_*$ and low-j$_*$ samples (not shown).  Together with Figure \ref{fig:MCGM_mapall}, this indicates that while on average the mass distribution of breakBRD galaxies is likely less extended than the comparison sample (Figure \ref{fig:MCGM_mapall}), there could be a large amount of overlap in the samples.  

\begin{figure}
    \includegraphics[scale=0.41]{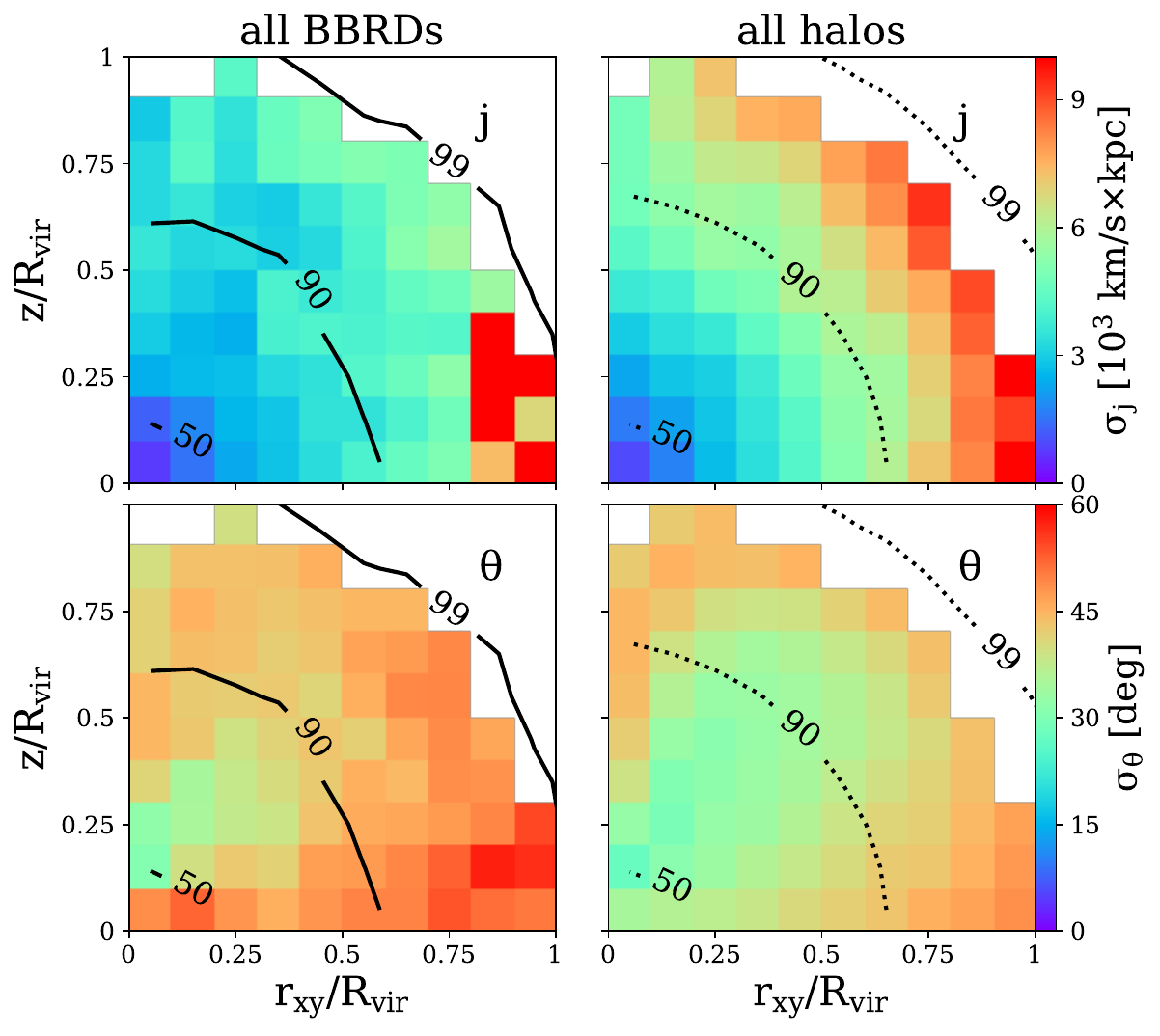}
    \caption{\textbf{Top:} The variation of the angular momentum distribution in the cold gas of the CGM between galaxies in both the breakBRD (left) and comparison (right) populations in each mapped cell.  \textbf{Bottom:} The variation of the misalignment angle of j$_*$ and j$_{\rm CGM}$ in the cold gas of the CGM between galaxies in both the breakBRD (left) and comparison (right) populations in each mapped cell.   } \label{fig:app_jbetweengals}
\end{figure}

\begin{figure}
    \includegraphics[scale=0.41]{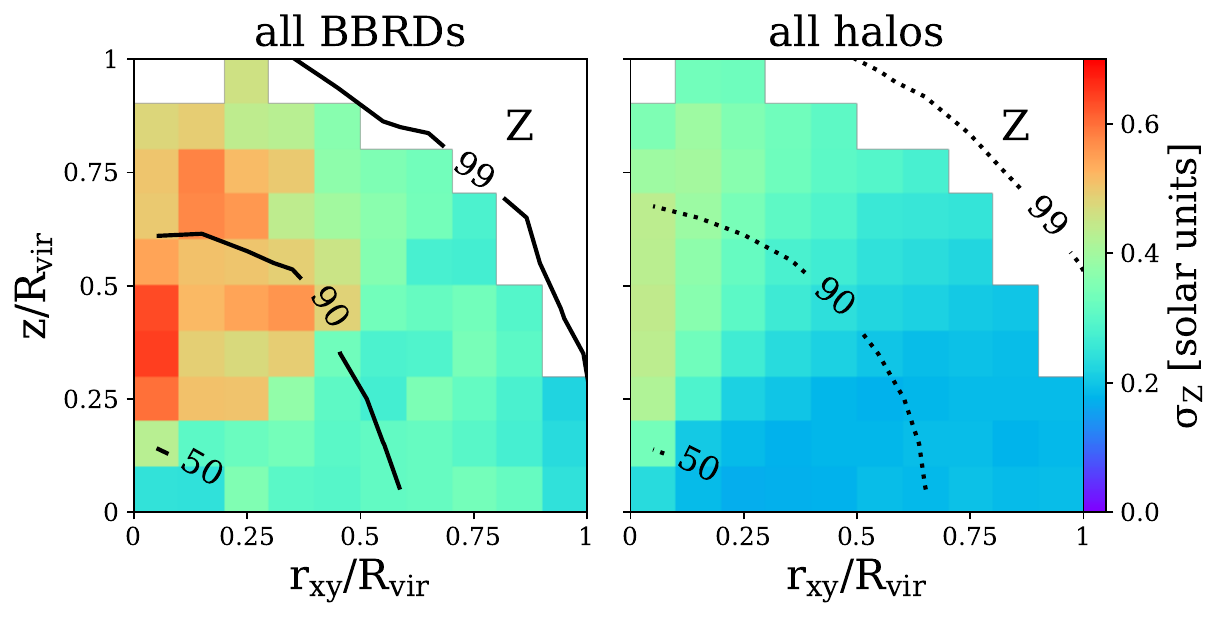}
    \caption{The variation of the metallicity distribution in the cold gas of the CGM between galaxies in both the breakBRD (left) and comparison (right) populations in each mapped cell.   } \label{fig:app_Zbetweengals}
\end{figure}

In Figure \ref{fig:app_jbetweengals} we map the variation of the cold gas angular momentum in the top panels and the variation in the misalignment angle of j$_*$ and j$_{\rm CGM}$ in the bottom panels.  In the top panels, we see that near the galaxy disk and closer to the galaxy center the angular momentum magnitude dispersion between galaxies is relatively small.  In the outskirts the dispersion between galaxies increases, where there is little CGM mass and the average j$_{\rm CGM}$ is slightly larger.  The breakBRD population shows somewhat lower $\sigma_{\rm j}$ values, particularly at larger radii.  The the bottom panels show a larger $\sigma_{\rm \theta}$ for the breakBRD galaxies, particularly along the disk plane, indicating that disk rotation is not as universal in the breakBRD sample.    

Finally, in Figure \ref{fig:app_Zbetweengals} we show maps of the metallicity dispersion between galaxies.  The dispersion in metallicity indicates that, as with the other CGM properties, although the average behavior of the CGM differs, there is still significant overlap in the metallicity values found in the CGM of breakBRD analogues and the comparison population of galaxies.  The high dispersion values near the polar region of the breakBRD sample is driven by the high metallicity values of high-j$_{*}$ breakBRD analogues.

\section{Testing Symmetry within the CGM}\label{app:CGMsymmetry}

As we discuss in Section \ref{sec:maps}, we assume that the CGM around galaxies is symmetric above and below the disk as well as azimuthally, and varies as a function of cylindrical radius and height above the disk.  We can test this symmetry assumption by finding the 1-$\sigma$ dispersion within galaxies at each mapped grid cell.  In this Appendix we show the maps of the average of these dispersion measures, giving an average dispersion within the galaxies for each of our samples (breakBRD and the comparison galaxies).  

\begin{figure}
    \includegraphics[scale=0.41]{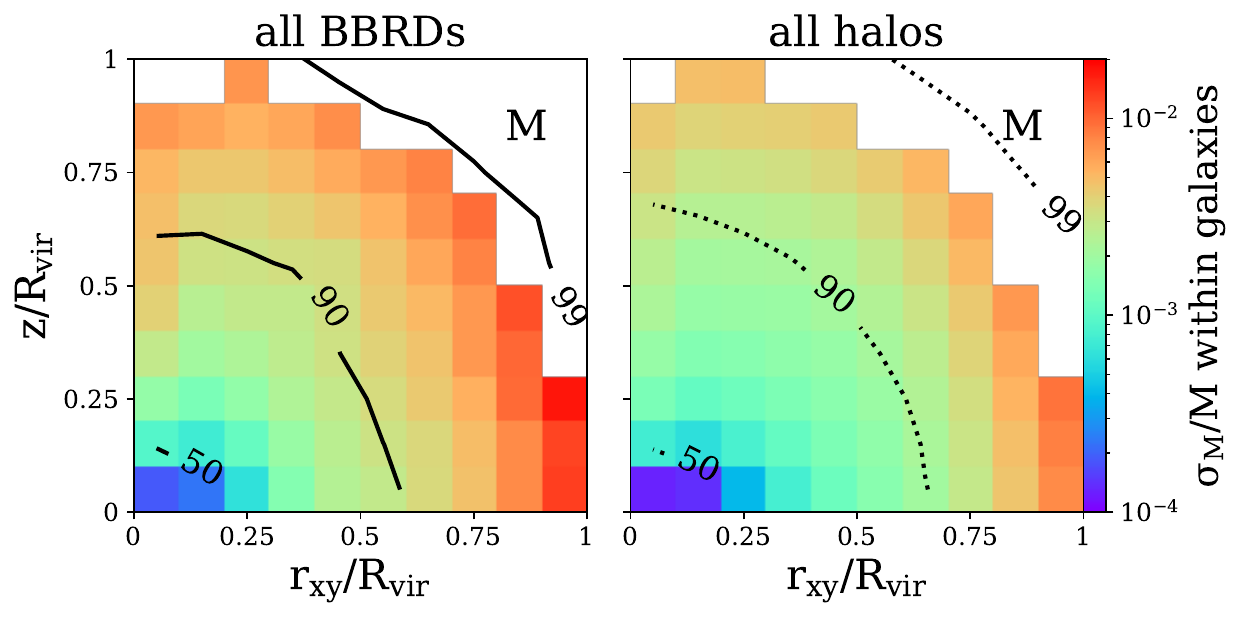}
    \caption{The average variation of the cold gas mass distribution in the CGM within galaxies in both the breakBRD (left) and comparison (right) populations relative to the mass in each mapped cell.   } \label{fig:app_masswithingals}
\end{figure}

\begin{figure}
    \includegraphics[scale=0.41]{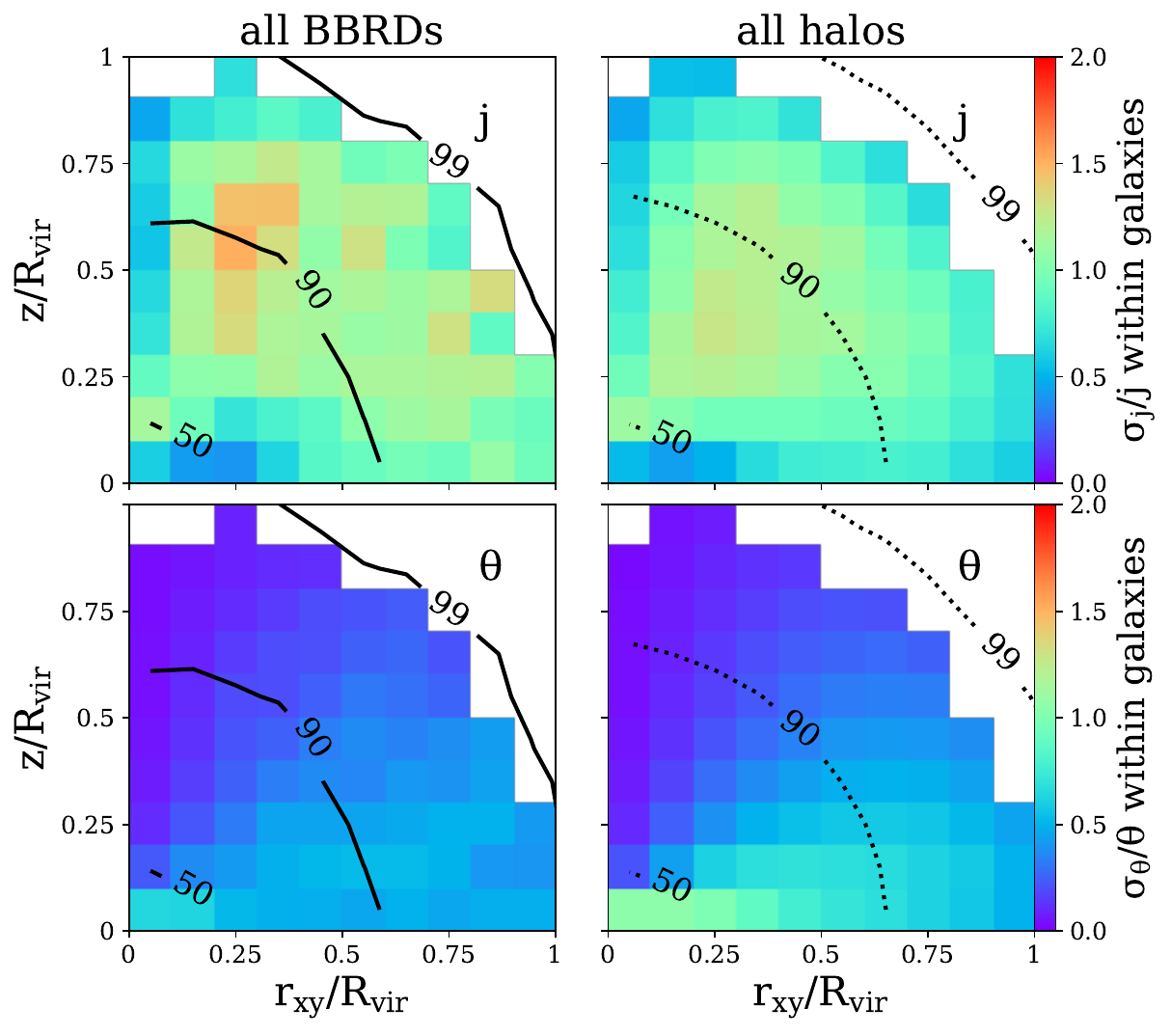}
    \caption{\textbf{Top:} The average variation of the angular momentum distribution in the cold gas of the CGM within galaxies in both the breakBRD (left) and comparison (right) populations relative to the angular momentum in each cell.  \textbf{Bottom:} The variation of the misalignment angle of j$_*$ and j$_{\rm CGM}$ in the cold gas of the CGM between galaxies in both the breakBRD (left) and comparison (right) populations relative to the misalignment in each mapped cell.   } \label{fig:app_jwithingals}
\end{figure}

\begin{figure}
    \includegraphics[scale=0.41]{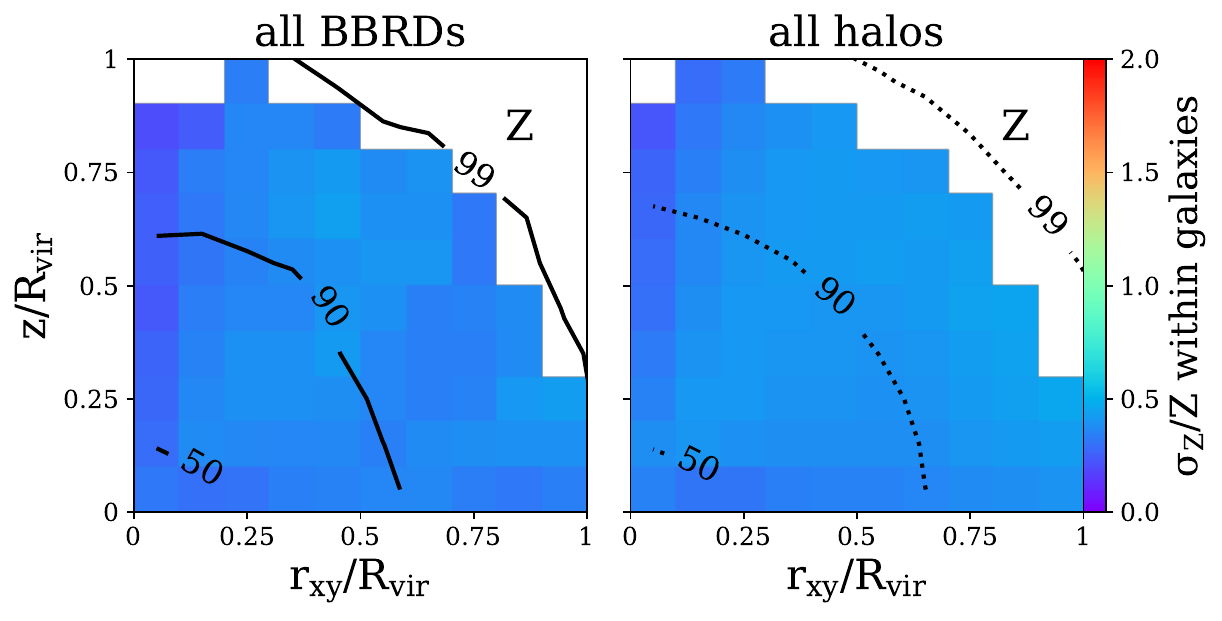}
    \caption{The average variation of the metallicity distribution in the cold gas of the CGM within galaxies in both the breakBRD (left) and comparison (right) populations relative to the metallicity in each mapped cell.   } \label{fig:app_Zwithingals}
\end{figure}

Briefly, the dispersion maps are similar for the breakBRD and the comparison population, indicating a similar level of (a)symmetry in the two samples for all of the CGM properties shown.  The mass (\ref{fig:app_masswithingals}) maps show dispersions of less than 2\% throughout the CGM, indicating that the cold gas is quite symmetrically distributed within these $\sim$30 kpc cells.  Both the angular momentum (\ref{fig:app_jwithingals}) and  metallicity (\ref{fig:app_Zwithingals}) maps show significantly more dispersion throughout the CGM. 
 The metallicity maps (\ref{fig:app_Zwithingals}) show relatively uniform dispersions throughout the CGM of about 40\%, while the angular momentum maps (\ref{fig:app_jwithingals}) show higher dispersion as one looks away from the pole or disk plane where the gas motions are less likely to be dominated by either outflows or disk rotation, respectively.  In the bottom panels of Figure \ref{fig:app_jwithingals}, the larger $\sigma_{\rm \theta}$/${\rm theta}$ values near the disk plane are more indicative of the small misalignment between j$_*$ and j$_{\rm CGM}$ (as shown in Figure \ref{fig:jCGM_mapall}) than a large $\sigma_{\rm \theta}$.  

\end{appendix}

\end{document}